\documentclass[a4paper,12pt]{article}
\usepackage[utf8]{inputenc}
\newcommand{\DATUM}{08-Nov-2022}
\usepackage{amsmath}  
\usepackage{amssymb}       
\usepackage{amsfonts}
\usepackage{bbm}
\usepackage{times}
\usepackage{amsthm} 
\usepackage{epsfig}
\usepackage{paralist}         
\usepackage{color}
\newcommand{\ol}{\overline}   
  
\newcommand{\eps}{{\varepsilon}}  
\newcommand{\vphi}{{\varphi}}           
\newcommand{\Om}{\Omega}                
\newcommand{\om}{\omega}
\newcommand{\la}{\langle}
\newcommand{\ra}{\rangle}
\newcommand{\udarrow}{\uparrow, \downarrow}
\newcommand{\bfone}{\mathbf{1}}
\newcommand{\cB}{\mathcal{B}}

\newcommand{\cE}{\mathcal{E}}
\newcommand{\cF}{\mathcal{F}}

\newcommand{\cI}{\mathcal{I}}

\newcommand{\cL}{\mathcal{L}}

\newcommand{\cS}{\mathcal{S}}

\newcommand{\cU}{\mathcal{U}}

\newcommand{\bbA}{\mathbbm{A}} 
 
\newcommand{\BB}{\mathbbm{B}} 
\newcommand{\CC}{\mathbbm{C}}     
\newcommand{\EE}{\mathbbm{E}} 
\newcommand{\HH}{\mathbbm{H}} 
\newcommand{\tHH}{{\widetilde{\mathbbm{H}}}} 
     
\newcommand{\NN}{\mathbbm{N}}     
\newcommand{\PP}{\mathbbm{P}}

\newcommand{\QQ}{\mathbbm{Q}}     
\newcommand{\RR}{\mathbbm{R}}     
\newcommand{\UU}{\mathbbm{U}} 
 
\newcommand{\TT}{\mathbbm{T}}     
\newcommand{\VV}{\mathbbm{V}}     
     
\newcommand{\ZZ}{\mathbbm{Z}}     
\newcommand{\fF}{\mathfrak{F}}

\newcommand{\fh}{\mathfrak{h}}

\newcommand{\fg}{\mathfrak{g}} 
\newcommand{\fs}{\mathfrak{s}}  
\newcommand{\sfj}{\mathsf{j}}

\newcommand{\uw}{{\underline w}}

\newcommand{\hc}{\hat{c}}

\newcommand{\hatom}{\hat{\om}}

\newcommand{\tLambda}{\widetilde{\Lambda}}

\newcommand{\vkappa}{{\vec{\kappa}}}
\newcommand{\vsigma}{{\vec{\sigma}}}

\newcommand{\va}{{\vec{a}}}
\newcommand{\vb}{{\vec{b}}}
\newcommand{\ve}{{\vec{e}}}

\newcommand{\vq}{{\vec{q}}}
\newcommand{\vv}{{\vec{v}}}
\newcommand{\vw}{{\vec{w}}}

\newcommand{\vz}{{\vec{z}}}

\newcommand{\fin}{\mathrm{fin}}

\newcommand{\rRe}{\mathrm{Re}}               
              
\newcommand{\Ran}{\mathrm{Ran}}              
\newcommand{\cirS}{\mathop{\bigcirc\kern -.73em {\scriptstyle{\mathrm{S}}}}}

\newcommand{\huelle}{{\mathrm{span}}}

\newcommand{\Tr}{{\mathrm{Tr}}}

\newcommand{\op}{{\mathrm{op}}} 
\newcommand{\gs}{{\mathrm{gs}}}
\newcommand{\HF}{{\scriptscriptstyle \mathrm{HF}}}
\newcommand{\eHF}{e^{\scriptscriptstyle (\mathrm{HF})}}
\newcommand{\fHF}{f^{\scriptscriptstyle (\mathrm{HF})}}

\newcommand{\Ex}{{\mathrm{Ex}}}

\newcommand{\main}{{\mathrm{main}}}
\newcommand{\rem}{{\mathrm{rem}}}

\newcommand{\mun}{{\mathrm{min}}}
\newcommand{\cIh}{{\cI(h)}}
\newcommand{\cIl}{{\cI(\ell)}}
\newcommand{\secction}[1]{\section{#1}\setcounter{equation}{0}}

\newtheorem{theorem}{Theorem}[section]

\theoremstyle{plain}
\begin{document}
\bibliographystyle{plain}
\title{On Relative Bounds for Interacting Fermion Operators}

\author{Volker~Bach \thanks{Institut f\"ur Analysis und Algebra,
    Technische Universit\"at Braunschweig, Germany,
    $<$v.bach@tu-bs.de$>$, ORCID 0000-0003-3987-8155} 
\and Robert Rauch \thanks{Institut f\"ur Analysis und Algebra,
    Technische Universit\"at Braunschweig, Germany,
    $<$r.rauch@tu-bs.de$>$, ORCID 0000-0003-4421-1216}
}

\date{\DATUM}

\maketitle

\begin{abstract} We consider a Hubbard model with nearest neighbor
  interaction on a discrete $d$-dimensional torus of length $L$ around
  its Hartree-Fock ground state and derive relative bounds of the
  effective interaction with respect to the effective kinetic
  energy. It is shown that there are no relative bounds uniform in
  $L$.
\end{abstract}

\textbf{MSC}: 81-02, 81Q05, 81V45, 81V55, 81V74
  
\textbf{Keywords}: Relative Bounds $\cdot$ Perturbation Theory $\cdot$
Coulomb Systems

\newpage
\secction{Introduction and Main Result} 
\label{sec-I}
%
All models of matter in physics and chemistry used in science and
technology ultimately derive from the quantum mechanical description
of interacting many-body systems. The precise description of these
interacting quantum many-body systems is one of the most important
tasks of mathematical and theoretical physics. The conceptual and
mathematical framework was formulated almost a century ago and has not
changed much since. Yet, the analysis especially of interacting
systems is complex and remains challenging.

In this paper, we consider a many-fermion quantum system whose states are
represented by vectors in a fermion Fock space
$\fF=\fF_f(\fh)$, where $\fh$ is the Hilbert
space of a single fermion, and a second-quantized Hamiltonian
\begin{align} \label{eq-Ia.01}
\tHH \ = \ \TT \: + \: \frac{g}{2} \, \VV
\end{align}
acting on a suitable domain in $\fF$. Here, $\TT$ is a one-particle
operator which is quadratic in the fields and represents the kinetic
energy and external fields, $\VV \geq 0$ is the purely repulsive pair
interaction between the fermions and quartic in the fields, and $g>0$
is a small coupling constant. This is the standard framework which is
described with mathematical precision, e.g., in \cite{ThirringIII1979,
  ThirringIV1980, ReedSimonII1980, ReedSimonIV1978,
  BratteliRobinson-II-1996}.

Assume the $N$-fermion Slater determinant 
$\Phi_\HF^{(N)} = \fHF_1 \wedge \cdots \wedge \fHF_N$ to be a
Hartree--Fock ground state. It induces a unitary particle-hole
Bogoliubov transformation $\UU_\HF$ on $\fF$. After Wick-ordering,
the transformed Hamiltonian assumes the form
\begin{align} \label{eq-Ia.02}
\HH \ := \ 
\UU_\HF^* \, \tHH \, \UU_\HF
\ = \ 
E_\HF^{(N)} \: + \: \TT_\HF \: + \: \frac{g}{2} \, \QQ \, ,
\end{align}
where the constant $E_\HF^{(N)} = \la \Om | \HH \Om \ra$ is the vacuum
expectation value of the transformed Hamiltonian, $\TT_\HF$ is
quadratic and normal-ordered in the field operators, and $\QQ$ is
quartic and normal-ordered in the field operators. More specifically,
it turns out that 
$E_\HF^{(N)} := \la \Phi_\HF^{(N)} | \tHH \Phi_\HF^{(N)} \ra$ is the
Hartree--Fock energy of the system and that $\TT_\HF \geq 0$, 
called the Hartree--Fock Hamiltonian, is the second quantization of a
positive effective one-body operator. We think of \eqref{eq-Ia.02} 
as an expansion of $\tHH$ around the transformed Hartree--Fock ground state
$\Phi_{\HF}^{(N)}$, where $\QQ$ encodes the properties of the
system beyond Hartree--Fock theory. For a review of Hartree--Fock
theory we refer the reader to \cite{Bach2022a}.

To develop a rigorous perturbation theory for the many-fermion system
in an operator-theoretic framework, it is natural to decompose $\QQ$
as $\QQ = \QQ_\main + \QQ_\rem$, where $\QQ_\main \geq 0$ and
$\QQ_\rem$ is relatively bounded by $\TT_\HF + \frac{g}{2} \QQ_\main$
with a small relative bound. The main result of this paper is
that this idea fails and that, in general, there is no such
decomposition. We demonstrate this statement by constructing a
counterexample in several steps:
\begin{itemize}
\item[(i)] In Eqs.~\eqref{eq-III.14}-\eqref{eq-III.17}, we decompose
  $\QQ$ as $\QQ = \rRe[ \QQ_1 + \QQ_2 - 2 \QQ_3 + 2 \QQ_4 + 4 \QQ_5 +
  4 \QQ_6 + 2 \QQ_7]$, where $\QQ_\main = \QQ_1 + \QQ_2$ and $\QQ_1
  \geq 0$ and $\QQ_2 \geq 0$ is the particle-particle and the
  hole-hole repulsion, respectively.

\item[(ii)] For system of electrons (spin-$\frac{1}{2}$ fermions) on a
  periodic $d$-dimensional lattice $\Lambda = \ZZ_L^d$ of sidelength
  $L \in \ZZ^+$ with an interaction given by a repulsive pair
  potential $v: \Lambda \to \RR_0^+$ we show in
  Theorem~\ref{thm-III.01} that the quadratic forms corresponding to
  $\QQ_3$, $\QQ_4$, $\QQ_5$, and $\QQ_6$ are bounded relative to $\NN
  + \QQ_\main$, uniformly in the thermodynamic (TD) limit, i.e., as $L
  \to \infty$. Note that $\NN + \QQ_\main$ is comparable to $\HH_0 =
  \TT_\HF + \QQ_\main$, provided the effective one-body operator
  entering the Hartree--Fock Hamiltonian $\TT_\HF$ is strictly
  positive.

\item[(iii)] Our counterexample is built on $\QQ_7$ which is a sum of
  products of four creation operators, namely, two particle and two
  hole creation operators. Our first main result is
  Theorem~\ref{thm-III.02}, in which we define a normalized trial
  vector $\Phi_\eps = \sqrt{1-\eps^2} \Om + \eps \|\QQ_7\Om\|^{-1}
  \QQ_7\Om$, for $\eps \in (0,\frac{1}{2}]$. 
  We show that 
  $0 \leq \la \Phi_\eps | \TT_\HF \Phi_\eps \ra \leq 4 \|t\|_\op$ and
  $0 \leq \la \Phi_\eps | \frac{g}{2} \QQ_\main \Phi_\eps \ra \leq 2g
  \eps^2 \|v\|_\op$ are uniformly bounded in the TD limit, provided
  that the one-particle kinetic energy $t$ and the pair interaction
  $v$ are bounded.
  In contrast, $\la \Phi_\eps| \QQ_7 \Phi_\eps \ra = \frac{g}{2} \|
  \QQ_7 \Om \|$, and $\| \QQ_7 \Om \|^2$ is characterized in
  Theorem~\ref{thm-III.02} by \eqref{eq-III.48,05}, which suggests that
  $\| \QQ_7 \Om \|^2 \sim |\Lambda| = L^d$ is an extensive quantity, at
  least for translation invariant systems.

  Note that the importance of the term $\QQ_7$ in the perturbative
  expansions for fermion systems (and also for boson systems) has been
  observed before in \cite{BenedikterNamPortaSchleinSeiringer2020,
    HainzlPortaRexze2020,
    BenedikterNamPortaSchleinSeiringer2021}. These go beyond the
  results of the present paper in as much as unitary operators that
  approximately eliminate $\QQ_7$ have been
  constructed and proven to yield the next correction, e.g., in an
  expansion of the ground state energy in powers of the coupling
  constant.

\item[(iv)] In Theorem~\ref{thm-III.03} we choose a specific model
  which falls into the category of models considered in \textit{(ii)}
  and \textit{(iii)} above, namely, the Hubbard model at
  half-filling. Following the lines of \cite{BachLiebSolovej1994}
  for this model, $E_\HF^{(N)}$, $\TT_\HF$, and $\QQ$ can be
  explicitly computed. While any operator is relatively bounded to any
  other operator in case of finite-dimensional Hilbert spaces, we show
  here that $\QQ$ contains indefinite contributions $\QQ_\rem$
  which cannot be bounded relative to $\TT_\HF + \QQ_\main$ with a
  relative bound that is \textit{uniform in the thermodynamic
    (large-volume) limit}.
\end{itemize}

\secction{$N$-Fermion Systems and Hartree--Fock Approximation} 
\label{sec-II}
%
\paragraph*{$\boldsymbol{N}$-Fermion Systems.} 
The state of a system of $N \in \ZZ^+ := \{ 1, 2, 3, \ldots \}$
interacting nonrelativistic fermions at time $t \in \RR$ in an atom, a
molecule, or a crystal is described by a wave function $\Psi_t \in
\fF^{(N)} \equiv \fF^{(N)}(\fh)$ in, or more generally a density
matrix $\rho_t \in \cL_+^1(\fF^{(N)})$ on, the $N$-fermion Hilbert
space $\fF^{(N)}(\fh) \subseteq \fh^{\otimes N}$, which is the subspace
of totally antisymmetric vectors in the $N$-fold tensor product of the
one-particle Hilbert space $\fh$. The Hilbert space $\fF^{(N)}(\fh)$
is the closure of the span of $N$-fermion Slater determinants $f_1
\wedge \cdots \wedge f_N \: := \: (N!)^{-1/2} \sum_{\pi \in \cS_N}
(-1)^\pi \: f_{\pi(1)} \otimes \cdots \otimes f_{\pi(N)}$. Here,
$\cS_N$ is the set of permutations of $\{1, \ldots, N\}$ and
$(-1)^\pi$ denotes their sign. If $\{ f_k \}_{k=1}^D \subseteq \fh$ is
an orthonormal basis (ONB) then so is $\big\{ f_{k(1)} \wedge \cdots
\wedge f_{k(N)} \: \big| 1 \leq k(1) < \ldots < k(N) \leq D \big\}
\subseteq \fF^{(N)}(\fh)$, where $D := \dim(\fh) \in \ZZ^+ \cup \{
\infty \}$ is the dimension of the one-particle Hilbert space.

The dynamics of the $N$-fermion system is determined by the
Schr{\"o}dinger equation $i \dot{\Psi}_t = H^{(N)} \Psi_t$ or 
$i \dot{\rho}_t = [H^{(N)}, \rho_t]$, respectively. Its generator is
the self-adjoint Hamiltonian
\begin{align} \label{eq-II.01}
H^{(N)} \ := \ T^{(N)} & \: + \: \frac{\kappa}{2} \, V^{(N)} \, , 
\quad \text{where}
\\[1ex]  \label{eq-II.02}
T^{(N)} \ := \ \sum_{m=1}^N T_m 
\, , & \quad 
V^{(N)} \ := \ \sum_{m,n = 1; m \neq n}^N V_{m,n} \, ,
\end{align}
$T_m$ is the one-body Hamiltonian $t$ acting on the $m^{th}$ variable,
$V_{m,n}$ is the pair interaction $v$ acting on the $m^{th}$ and the
$n^{th}$ variable, and $\kappa > 0$ is a coupling constant. That is,
$T_m = \Pi_m^* \circ (t \otimes \bfone \otimes \cdots \otimes \bfone)
\circ \Pi_m$ and $V_{m,n} = \Pi_{m,n}^* \circ (v \otimes \bfone
\otimes \cdots \otimes \bfone) \circ \Pi_{m,n}$, where $\Pi_m$ 
is the natural permutation operator exchanging the factor $f_1$ with
$f_m$, and $\Pi_{m,n}$ is the natural permutation operator exchanging
the factors $f_1$ with $f_m$ and $f_2$ with $f_n$, in the tensor
product $f_1 \otimes f_2 \otimes \cdots \otimes f_N$.

The semiboundedness and self-adjointness of $H^{(N)}$ can be ensured
by the assumption that $t: \fs \to \fh$ is a semibounded and
self-adjoint linear operator defined on a dense domain $\fs \subseteq
\fh$ and that $v: \fs \otimes \fs \to \fh \otimes \fh$ is a symmetric,
nonnegative linear operator and an infinitesimal perturbation of $t
\otimes \bfone + \bfone \otimes t$. Furthermore, we assume
w.l.o.g.\ that $v$ is invariant under exchanging the tensor factors in
$\fh \otimes \fh$, i.e., that $\Ex \circ v = v \circ \Ex$, where the
\textit{exchange operator} $\Ex \in \cB[\fh \otimes \fh]$ is defined
by $\Ex( f \otimes g) = g \otimes f$. Then $H^{(N)}$ is semibounded
and essentially self-adjoint on the subspace 
$\fF_\fin^{(N)}(\fs) \subseteq \fF^{(N)}(\fh)$ of (finite) linear
combinations of Slater determinants $f_1 \wedge \cdots \wedge f_N$
with $f_1, \ldots, f_N \in \fs$.

Basic quantities for the description of an $N$-fermion system are its
ground state energy $E_\gs^{(N)}$ defined to the smallest expectation
value of $H^{(N)}$ evaluated on $N$-fermion wave functions,
\begin{align} \label{eq-II.02,01}
E_\gs^{(N)} \ := \ 
\inf\Big\{ \la \Psi^{(N)} \, | \; H^{(N)} \, \Psi^{(N)} \ra \; \Big| 
\ \Psi^{(N)} \in \fF^{(N)}(\fh) \cap \fs^{\otimes N} \, , \ 
\|\Psi^{(N)}\| = 1 \; \Big\} \, , 
\end{align}
and, if existent, the corresponding minimizers $\Psi_\gs^{(N)} \in 
\fF^{(N)}(\fh) \cap \fs^{\otimes N}$ called ground states, which
necessarily fulfill the time-independent Schrödinger equation
$H^{(N)} \Psi^{(N)} = E_\gs^{(N)} \Psi^{(N)}$.

\paragraph*{Fock Space, CAR, and Second Quantization.} 
It is convenient to consider $\fF^{(N)}(\fh)$ a subspace of the
fermion Fock space $\fF(\fh) = \bigoplus_{N=0}^\infty \fF^{(N)}(\fh)$,
where $\fF^{(0)} = \CC \cdot \Om$ is the one-dimensional vacuum
subspace spanned by the normalized vacuum vector $\Om$. We introduce
the usual fermion creation operators $\{ c^*(f) \}_{f \in \fh}
\subseteq \cB[\fF]$ as follows. For a fixed orbital $f \in \fh$ and $N
< D$, these are bounded operators $c^*(f) \in \cB[\fF^{(N)};
  \fF^{(N+1)}]$ defined by their action $c^*(f) \Om := f$ on the
vacuum vector, for $N = 0$, and $c^*(f) [g_1 \wedge \cdots \wedge g_N]
:= f \wedge g_1 \wedge \cdots \wedge g_N$ on Slater determinants, for
$N \in \ZZ^+$ and $g_1, \ldots, g_N \in \fh$. Extending these
definitions by linearity and continuity to all of $\fF$, one obtains a
family $\{ c^*(f) \}_{f \in \fh} \subseteq \cB[\fF]$ of bounded
operators on $\fF$ whose norm equals $\| c^*(f) \| = \| f \|$. The
Slater determinants can now be rewritten as $f_1 \wedge \cdots \wedge
f_N = c^*(f_1) \cdots c^*(f_N) \Om$, and from an ONB $\{ f_k
\}_{k=1}^D \subseteq \fh$ of the one-particle Hilbert space we obtain
ONB
\begin{align} \label{eq-II.03}
\big\{ c^*(f_{k(1)}) \cdots c^*(f_{k(N)}) \Om \: \big| 
\ 1 \leq k(1) < \ldots < k(N) \leq D \big\} 
& \ \subseteq \ \fF^{(N)}(\fh) \, ,
\\[1ex]
 \label{eq-II.04}
\bigcup_{N=0}^\infty \big\{ c^*(f_{k(1)}) \cdots c^*(f_{k(N)}) \Om \: \big| 
\ 1 \leq k(1) < \ldots < k(N) \leq D \big\} 
& \ \subseteq \ \fF(\fh) \, ,
\end{align}
of the $N$-fermion Hilbert space and the fermion Fock space,
respectively.

Given an orbital $f \in \fh$, the adjoint $c(f) := [c^*(f)]^* \in
\cB[\fF]$ of the creation operator $c^*(f)$ is called annihilation
operator. Creation and annihilation operators $\{ c^*(f), c(f) \}_{f
  \in \fh}$ form a Fock representation of the canonical
anticommutation relations (CAR), i.e., for all $f,g \in \fh$,
\begin{align} \label{eq-II.05}
\{ c^*(f), c^*(g) \} \ = \ \{ c(f), c(g) \} \ = \ 0 \, , \quad
\{ c(f), c^*(g) \} \ = \ \la f | g \ra \cdot \bfone_\fF \, , \quad
c(f) \Om \ = \ 0 \, .
\end{align}
We introduce the number operator $\NN$ and the second quantizations of
$H^{(N)} = T^{(N)} + \frac{\kappa}{2} V^{(N)}$, as defined in
\eqref{eq-II.01}, and its constituents $T^{(N)}$ and $V^{(N)}$ by
\begin{align} \label{eq-II.06}
\NN \ := \ \bigoplus_{N=0}^\infty N \, , \quad 
\TT & \ := \ \bigoplus_{N=0}^\infty T^{(N)} \, , \quad 
\VV \ := \ \bigoplus_{N=0}^\infty V^{(N)} \, , \quad
\\[1ex] \label{eq-II.06,1}
\tHH \ := \ & \bigoplus_{N=0}^\infty H^{(N)} 
\ = \ \TT + \frac{\kappa}{2} \VV \, .
\end{align}
These operators are essentially self-adjoint on the subspace
$\fF_\fin(\fs) \subseteq \fF(\fh)$ of finite vectors, i.e., finite
linear combinations of Slater determinants $f_1 \wedge \cdots \wedge
f_N$ with $f_1, \ldots, f_N \in \fs$ and varying $N \in \ZZ_0^+$.
Using an ONB $\{ f_k \}_{k \in \cI} \subseteq \fs$ of orbitals in
$\fh$, where $\cI := \{1, 2, \ldots, D\}$, the number operator and the
second quantized operators $\TT$ and $\VV$ -and hence also $\tHH$- can
be represented as $\NN = \sum_{k \in \cI} c^*(f_k) c(f_k)$,
\begin{align} \label{eq-II.07}
\TT \ = \ &
\sum_{k, m \in \cI} \la f_k \, | \; t \, f_m \ra 
\: c^*(f_k) \, c(f_m) \, , \quad 
\\[1ex] \label{eq-II.08}
\VV \ = \ &
\sum_{k, \ell, m, n \in \cI} 
\la f_k \otimes f_\ell \, | \; v \, (f_m \otimes f_n) \ra \: 
c^*(f_\ell) \, c^*(f_k) \, c(f_m) \, c(f_n) \, .
\end{align}
In case of unbounded $t$ or $v$, the existence of the matrix elements
$\la f_k | t \, f_m \ra$ and 
$\la f_k \otimes f_\ell | v (f_m \otimes f_n) \ra$ is guaranteed by
sufficient regularity of the elements of $\fs$.

\paragraph*{Finite Dimension.} 
For the purpose of this paper, the unboundedness of the operators
under consideration is an unnecessary complication, and we hence
simply assume that the dimension
\begin{align} \label{eq-II.09}
D \ = \ \dim(\fh) \ < \ \infty 
\end{align}
of the one-particle Hilbert space $\fh$ is finite and that $D > N$,
where the latter requirement ensures that statements we make are not
void. Consequently, the Fock space $\fF(\fh)$ is finite-dimensional,
too, namely, $\dim[\fF(\fh)] = 2^D < \infty$. Thanks to
Assumption~\eqref{eq-II.09}, the linear operators $t$, $v$, $\NN$,
$\TT$, $\VV$, and $\tHH$ are actually all finite-dimensional
self-adjoint matrices, $\fs = \fh$ and $\fF_\fin(\fs) = \fF(\fh)$. The
description of the theory without the assumption of finite dimension
can be found, e.g., in \cite{Bach2022a}. In the end, the assertions
formulated in our theorems become non-trivial in the asympotic limit
$D > N \gg 1$.

\paragraph*{Hartree--Fock Approximation and Bogoliubov Transformations.} 
The computation of the ground state energy $E_\gs^{(N)}$ and the
corresponding ground state(s) $\Psi_\gs^{(N)}$ is far too complicated,
due to the large dimension of the problem, even though the finiteness
of $D$ ensures their existence. The Hartree--Fock approximation
described below is one of the most important methods for $N$-fermion
systems.

The Hartree--Fock energy $E_\HF^{(N)}$ is defined to be the smallest
expectation value of $\tHH$ evaluated on $N$-fermion Slater
determinants,
\begin{align} \label{eq-II.10}
E_\HF^{(N)} \ := \ 
\inf\Big\{ \la f_1 \wedge \cdots \wedge f_N \, | \; \tHH \, 
f_1 \wedge \cdots \wedge f_N \ra \; \Big| \ f_j \in \fs \, , \ 
\la f_i | f_j \ra = \delta_{i,j} \; \Big\} \, . 
\end{align}
Note that, for orthonormal $f_1, \ldots, f_N \in \fs$, 
\begin{align} \label{eq-II.10,01}
\la f_1 \wedge \cdots & \wedge f_N \, | \; \tHH \, 
f_1 \wedge \cdots \wedge f_N \ra 
\ = \ 
\\[1ex] \nonumber
& \cE_\HF(\gamma) 
\ := \
\Tr_\fh[t \, \gamma] \: + \: 
\frac{\kappa}{2} \Tr_{\fh \otimes \fh}\big[ v \, 
(\bfone - \Ex) ( \gamma \otimes \gamma) \big] \, ,
\end{align}
where $\gamma = \sum_{\nu = 1}^N |f_\nu \ra\la f_\nu| = \gamma^* = \gamma^2$
is the rank-$N$ orthogonal projection onto the linear span of
the orbitals $f_1, \ldots, f_N$, and $\Ex \in \cB(\fh)$ is the exchange 
operator determined by $\Ex(f \otimes g) = g \otimes f$. Therefore, 
\begin{align} \label{eq-II.10,02}
E_\HF^{(N)} \ = \ & 
\inf\Big\{ \cE_\HF(\gamma) \; \Big| \ \gamma = \gamma^* = \gamma^2 \, ,
\ \Tr(\gamma) = N \; \Big\} 
\nonumber \\[1ex] 
\ = \ & 
\inf\Big\{ \cE_\HF(\gamma) \; \Big| \ 0 \leq \gamma \leq \bfone \, ,
\ \Tr(\gamma) = N \; \Big\} \, ,
\end{align}
where the second equality is known as Lieb's variational principle
\cite{Lieb1981a, Bach1992}. Note that $\{ \gamma \in \cL^1(\fh) |
0 \leq \gamma \leq \bfone, \, \Tr(\gamma) = 1 \} \subseteq \cL^1(\fh)$
is a closed convex subset.

Thanks to $D < \infty$, the infimum in \eqref{eq-II.10,02} is actually
always a minimum attained at $P_\HF = \sum_{\nu = 1}^N
|\fHF_\nu \ra\la \fHF_\nu|$, say, with orthonormal $\{\fHF_1,
\ldots, \fHF_N \} \subseteq \fh$. The minimizer(s) $P_\HF$,
called the Hartree--Fock ground state, fulfills a stationarity
condition
\begin{align} \label{eq-II.10,03}
P_\HF \ = \ &
\bfone_N \big[ h_\HF(P_\HF) \big] \, , 
\end{align}
known as the Hartree--Fock equation, where $\bfone_N$ denotes the
projection onto the lowest $N$ eigenvalues (counting multiplicities)
of the Hartree--Fock effective Hamiltonian $h_\HF(P_\HF) \in
\cB[\fh]$, which is determined by
\begin{align} \label{eq-II.11}
\Tr[ h_\HF(\gamma) \, \eta ]
\ = \
\Tr_\fh[t \, \eta] \: + \: 
\kappa \, \Tr_{\fh \otimes \fh}\big[ v \, 
(\bfone - \Ex) ( \gamma \otimes \eta) \big] \, ,
\end{align}
for all trace-class operators $\eta \in \cL^1(\fh)$.
Assuming w.l.o.g.\ that the eigenvalues $\eHF_j \in \RR$ of 
$h_\HF(P_\HF)$ are given in ascending order, 
$\eHF_1 \leq \eHF_2 \leq \ldots \leq \eHF_D$, we obtain an
ONB $\{\fHF_1, \ldots, \fHF_D \} \subseteq \fh$ of eigenvectors
of $h_\HF(P_\HF)$ with the first $N$ vectors being the orbitals that 
enter the Hartree--Fock ground state $P_\HF$. The \textit{no unfilled
  shells theorem} of (unrestricted) Hartree--Fock theory
\cite{BachLiebLossSolovej1994, BachLiebSolovej1994} ensures that
\begin{align} \label{eq-II.11,01}
\eHF_N \ < \ \mu_N 
\ := \ 
\frac{\eHF_N + \eHF_{N+1} }{2} 
\ < \ \eHF_{N+1} \, ,
\end{align}
and hence that there is no paradoxy in Eq.~\eqref{eq-II.10,03}
caused by $\dim\Ran\bfone[ h_\HF(P_\HF) \leq \eHF_N ] > N$.

\secction{Wick-Ordering and Relative Bounds} 
\label{sec-III}
%
\paragraph*{Wick-Ordering following a Bogoliubov Tranformation.} 
For each orbital $\fHF_k$ we abbreviate the corresponding creation
and annihilation operator by $c_k^* := c^*(\fHF_k)$ and $c_k :=
c(\fHF_k)$, respectively. Moreover, we define
\begin{align} \label{eq-III.01}
T_{k;m} \ := \ 
\la \fHF_k \, | \; t \, \fHF_m \ra 
\quad \text{and} \quad
V_{k,\ell;m,n} \ := \ 
\la \fHF_k \otimes \fHF_\ell \, | 
\; v \, (\fHF_m \otimes \fHF_n) \ra \, , 
\end{align}
such that
\begin{align} \label{eq-III.02}
\tHH \ = \ &
\sum_{k, m \in \cI} T_{k;m} \: c_k^* \, c_m 
\: + \: 
\frac{\kappa}{2} \sum_{k, \ell, m, n \in \cI} V_{k,\ell;m,n} 
\: c_\ell^* \, c_k^* \, c_m \, c_n \, .
\end{align}
Following the intiution that, for small $v$, the Hartree--Fock energy
$E_\HF^{(N)}$ and the Hartree--Fock ground state 
$\Phi_\HF := \fHF_1 \wedge \cdots \wedge \fHF_N \in \fF^{(N)}(\fh)$ 
are good approximations of the actual ground state energy
$E_\gs^{(N)}$ and a ground state $\Psi_\gs$, respectively, it is
natural to introduce a unitary operator $\UU_\HF \in \cU[\fF(\fh)]$ on
Fock space which transforms the vacuum vector $\Om$ into 
$\Phi_\HF = \UU_\HF \Om$, because then the Hartree--Fock energy
becomes the vacuum expectation value of $\tHH$ conjugated by the
unitary $\UU_\HF$,
\begin{align} \label{eq-III.03}
E_\HF^{(N)} \ = \ 
\big\la \Om \, \big| \: \UU_\HF^* \, \tHH \, \UU_\HF \, \Om \big\ra \, ,
\end{align}
as a natural offset for the energy. A unitary operator with this
property is the Bogoliubov
transformation $\UU_\HF$ defined by 
$\UU_\HF \Om := \Phi_\HF^{(N)} = c_1^* \cdots c_N^* \Om$ 
and
\begin{align} \label{eq-III.04}
\UU_\HF^* \, c^*(f) \, \UU_\HF
\ := \ & 
c^*\big( P_\HF^\perp f \big) \: + \: c\big( \sfj(P_\HF f) \big) \, ,
\end{align}
where 
\begin{align} \label{eq-III.04,1}
P_\HF \ := \ & \sum_{k=1}^N \big| \fHF_k \big\ra\big\la \fHF_k \big| 
\end{align}
and $\sfj: \fh \to \fh$ is the antiunitary involution defined by
$\sfj(\sum_{k = 1}^D \alpha_k \fHF_k) 
:= \sum_{k = 1}^D \ol{\alpha_k} \fHF_k$. 
Note that $P_\HF \circ \sfj = \sfj \circ P_\HF$.  It is convenient to
express this definition entirely in terms of the ONB $\{ \fHF_k
\}_{k=1}^D \subseteq \fh$ as
\begin{align} \label{eq-III.05}
\UU_\HF^* \, c_k^* \, \UU_\HF
\ := \ & 
h_k^* \: + \: \ell_k \, ,
\\[1ex] \label{eq-III.06}
h_k^* \ := \ \bfone_\cIh(k) \, c_k^* 
\quad & \text{and} \quad   
\ell_k \ := \ \bfone_\cIl(k) \, c_k \, , 
\end{align}
where $\cIh := \{ k \in \cI | k \geq N+1 \}$ and 
$\cIl := \{ k \in \cI | k \leq N \}$. The operators 
$\{h_k^*, h_k, \ell_k^*, \ell_k \}_{k \in \cI}$ are again a 
Fock representation of the CAR, i.e., for all $j, k \in \cI$,
\begin{align} \label{eq-III.07}
\{ h_k^* , h_j^* \} 
\ = \ &
\{ h_k , h_j \} 
\ = \ 
\{ \ell_k^* , \ell_j^* \} 
\ = \ 
\{ \ell_k , \ell_j \} 
\ = \ 
\{ h_k^* , \ell_j^* \} 
\ = \ 
\{ h_k , \ell_j \} 
\nonumber \\ 
\ = \ &
\{ h_k , \ell_j^* \} 
\ = \ 
\{ h_k^* , \ell_j \} 
\ = \ 0 \, , 
\\[1ex] \label{eq-III.08}
\{ h_k , h_j^* \} \ = \ & \delta_{k,j} \, \bfone_\cIh(k) \, , \ \ 
\{ \ell_k , \ell_j^* \} \ = \ \delta_{k,j} \, \bfone_\cIl(k) \, , \ \  
h_k \Om \ = \ \ell_k \Om \ = \ 0 \, ,
\end{align}
with respect to which the new number operator is
\begin{align} \label{eq-III.08,01}
\NN \ := \ \NN_h + \NN_\ell \, , 
\ \ \text{where} \ \ 
\NN_h \ := \ \sum_{k \in \cIh} h_k^* h_k \, ,   
\ \ 
\NN_\ell \ := \ \sum_{k \in \cIl} \ell_k^* \ell_k \, .   
\end{align}
Conjugating $\tHH$ with $\UU_\HF$, we obtain the transformed
Hamiltonian
\begin{align} \label{eq-III.09}
\HH \ := \ &
\UU_\HF^* \, \tHH \, \UU_\HF 
\ = \ 
\UU_\HF^* \, \TT \, \UU_\HF \: + \: 
\frac{\kappa}{2} \UU_\HF^* \, \VV \, \UU_\HF 
\nonumber \\[1ex] 
\ = \ &
\sum_{k, m \in \cI} T_{k;m} \, (h_k^* + \ell_k) (h_m + \ell_m^*) 
\\ \nonumber
& \: + \: 
\frac{\kappa}{2} \sum_{j, k, m, n \in \cI} V_{j, k; m,n} 
\, (h_k^* + \ell_k) (h_j^* + \ell_j) (h_m + \ell_m^*) (h_n + \ell_n^*) 
\, ,
\end{align}
and by Wick-ordering, i.e., anticommuting all creation operators
$h_k^*$ and $\ell_k^*$ to the left and all annihilation operators
$h_k$ and $\ell_k$ to the right, we rewrite the result
in the form
\begin{align} \label{eq-III.10}
\HH \ = \ &
E_\HF^{(N)} \: + \: \TT_\HF \: + \: \frac{\kappa}{2} \QQ \, ,
\end{align}
where the first term is indeed the Hartree--Fock energy,
\begin{align} \label{eq-III.11}
E_\HF^{(N)} \ = \ & 
\la \Phi_\HF | \HH \Phi_\HF \ra 
\\[1ex] \nonumber 
\ = \ &
\cE_\HF(P_\HF) \ = \ 
\Tr_\fh[t \, P_\HF] \: + \: 
\frac{\kappa}{2} 
\Tr_{\fh \otimes \fh}[v \, (\bfone - \Ex)(P_\HF \otimes P_\HF)] \, ,
\end{align}
and serves as an energy offset. The second term $\TT_\HF$ is the
second quantization of the positive one-particle operator
$|h_\HF(P_\HF) - \mu_N|$ and, hence, itself positive. It collects all
terms that are quadratic in the field operators and equals
\begin{align} \label{eq-III.12}
\TT_\HF \ = \ 
\sum_{k \in \cIh} \om_k \, h_k^* \, h_k 
\: + \:
\sum_{k \in \cIh} \om_k \, \ell_k^* \, \ell_k 
\ \geq \ 
\frac{1}{2} \, \om_\mun \, \NN 
\ \geq \ 0 \, ,
\end{align}
where 
\begin{align} \label{eq-III.13}
\om_k \ := \ |\eHF_k - \mu_N| 
\quad \text{and} \quad
\om_\mun \ := \ \min_{k \in \cI} \om_k \ > \ 0 \, , 
\end{align}
and we recall that $\mu_N \ = \ \frac{1}{2} (\eHF_{N+1} + \eHF_N)$.
Finally, the quartic terms in the Hamiltonian are
collected in 
$\QQ = \rRe[ \QQ_1 + \QQ_2 - 2 \QQ_3 + 2 \QQ_4 + 4 \QQ_5 + 4 \QQ_6 + 2 \QQ_7]$, 
with
\begin{align} \label{eq-III.14}
\QQ_1 \ := 
\sum_{j, k, m, n \in \cI} V_{j, k; m,n} \, h_k^* h_j^* h_m h_n 
\, , & \quad
\QQ_2 \ := 
\sum_{j, k, m, n \in \cI} V_{j, k; m,n} \, \ell_m^* \ell_n^* \ell_k \ell_j \, ,
\\ \label{eq-III.15}
\QQ_3 \ := 
\sum_{j, k, m, n \in \cI} V_{j, k; m,n} \, h_k^* \ell_m^* \ell_j h_n 
\, , & \quad
\QQ_4 \ := 
\sum_{j, k, m, n \in \cI} V_{j, k; m,n} \, h_j^* \ell_m^* \ell_k h_n  \, ,
\\ \label{eq-III.16}
\QQ_5 \ := 
\sum_{j, k, m, n \in \cI} V_{j, k; m,n} \, h_k^* \ell_m^* \ell_n^* \ell_j
\, , & \quad
\QQ_6 \ := 
\sum_{j, k, m, n \in \cI} V_{j, k; m,n} \, h_j^* h_m \ell_k h_n \, ,
\\ \label{eq-III.17}
\QQ_7 \ :=  
\sum_{j, k, m, n \in \cI} & V_{j, k; m,n} \, h_k^* h_j^* \ell_m^* \ell_n^* \, .
\end{align}

\paragraph*{Positivity of the Main Interaction Term
$\boldsymbol{\QQ_\main = \QQ_1 + \QQ_2}$.} 
We recall that the interaction potential $V \geq 0$ is assumed to be
positive. Hence, we may define $W := V^{1/2} \geq 0$ and observe that
$V_{j, k; m,n} = \sum_{p,q \in \cI} W_{j, k; p,q} W_{p,q; m,n}$. Introducing
\begin{align} \label{eq-III.18}
\bbA_{p,q} \ := \  \sum_{m, n \in \cI} W_{p,q; m,n} \, h_m h_n \, , \qquad 
\BB_{p,q} \ := \  \sum_{m, n \in \cI} W_{m,n; p,q} \, \ell_n \ell_m \, , 
\end{align}
we now observe that $\QQ_1$ and $\QQ_2$ are manifestly positive, 
\begin{align} \label{eq-III.19}
\QQ_1 \ = \ \sum_{p,q} \bbA_{p,q}^* \, \bbA_{p,q} \ \geq \ 0 \, , \qquad 
\QQ_2 \ = \ \sum_{p,q} \BB_{p,q}^* \, \BB_{p,q} \ \geq \ 0 \, .
\end{align}
Note that both absolute, but also relative, norm bounds on $\QQ_1$
become large as the dimension $D \gg N$ of the one-particle Hilbert
space $\fh$ growths large. The reason for this is that the number of
degrees of freedom corresponding to the transformed creation operators
$h_k^*$ is $D-N$. Since $\QQ_1$ is the only term in $\HH$ which
contains quartic terms in $h_k^*$ and $h_k$, i.e., monomials in
$h_k^*$ and $h_k$ of highest degree, it can never be relatively
bounded by the other terms in the Hamiltonian with a relative bound
which is uniform in $D \to \infty$. This fact holds true independent
of the regularity properties one may assume on the interaction
potential $v$. A similar argument applies to $\QQ_2$.

It is therefore natural to integrate the terms $\QQ_1$ and $\QQ_2$ in
what is considered the unperturbed Hamiltonian
\begin{align} \label{eq-III.20}
\HH_0 \ := \ \TT_\HF + \frac{\kappa}{2} \QQ_\main 
\quad \text{with} \quad
\QQ_\main \ := \ \QQ_1 + \QQ_2 \ \geq \ 0 \, , 
\end{align}
and treat the remaining sum $\QQ_\rem := \sum_{\nu=3}^7 \QQ_\nu$
as a perturbation of $\HH_0$,  
\begin{align} \label{eq-III.21}
\HH \ = \ \HH_0 + \frac{\kappa}{2} \QQ_\rem \, .
\end{align}
In fact, one would hope that the appearant big size of $\QQ_1$ and
$\QQ_2$ now turns into an advantage and helps to control the terms
entering $\QQ_\rem$. For this strategy to be successful, we need to
establish sufficiently strong bounds of $\QQ_\rem$ relative to
$\HH_0$, because then the spectral properties of $\HH$ could be
derived from those of $\HH_0$, provided the coupling constant $\kappa
> 0$ is sufficiently small. The main result of this paper, however, is
that this strategy fails, due to the presence of $\QQ_7$ in
$\QQ_\rem$, see \eqref{eq-III.17}.

\paragraph*{Smallness of the Interaction Terms
$\boldsymbol{\QQ_3, \QQ_4, \QQ_5}$ and $\boldsymbol{\QQ_6}$.} To
derive explicit bounds on the interaction terms $\QQ_\nu$, we specify
the model further and consider spin-$\frac{1}{2}$ particles on the
periodic $d$-dimensional lattice $\Lambda = \ZZ_L^d = (\ZZ/L\ZZ)^d$,
such that
\begin{align} \label{eq-III.22}
\fh \ = \ \ell^2\big( \Lambda \times \{\udarrow\} \big) 
\, , \qquad 
D \ = \ \dim(\fh) \ = \ 2 \, L^d \, .
\end{align}
The canonical ONB in $\fh$ is denoted $\{ \delta_{x,\sigma} \}_{x \in
  \Lambda, \sigma \in \{\udarrow\}} \subseteq \fh$, where
$\delta_{x,\sigma}(y,\tau) := \delta_{x,y} \delta_{\sigma,\tau}$. We
introduce the corresponding creation and annihilation operators by
$c_{x,\sigma}^* := c^*(\delta_{x,\sigma})$ and $c_{x,\sigma} =
c(\delta_{x,\sigma})$, for $x \in \Lambda$ and $\sigma \in \{ \udarrow
\}$. Hence, 
\begin{align} \label{eq-III.23}
\fF_f(\fh) \ = \ 
\huelle\big\{ c_{x_1,\sigma_1}^* \cdots c_{x_N,\sigma_N}^* \Om \: \big| 
\ N \in \NN_0 \, , \ \ x_i \in \Lambda \, , \ \ 
\sigma_i \in \{\udarrow\} \big\} \, .
\end{align}
The interaction $\VV$ in \eqref{eq-II.08} is assumed to be of the
usual form, i.e., to be induced by a nonnegative, spin-independent,
pair potential $v : \Lambda \to \RR_0^+$. It takes the familiar form
\begin{align} \label{eq-III.24}
\VV \ = \ &
\sum_{x,y \in \Lambda} \sum_{\sigma, \tau \in \{\udarrow\} } 
v_{x-y} \: c_{x,\sigma}^* \, c_{y,\tau}^* \, c_{y,\tau} \, c_{x,\sigma} \, .
\end{align}
Defining 
\begin{align} \label{eq-III.24,01}
h_{x,\sigma}^* \ := \ \sum_{k \in \cIh} \fHF_k(x,\sigma) \, c^*(\fHF_k))
\quad \text{and} \quad
\ell_{x,\sigma}^* \ := \ \sum_{k \in \cIl} \fHF_k(x,\sigma) \, c^*(\fHF_k)) \, ,
\end{align}
we observe that 
\begin{align} \label{eq-III.25}
\{ h_{x,\sigma}^* , h_{y,\tau}^* \} 
\ = \ &
\{ h_{x,\sigma} , h_{y,\tau} \} 
\ = \ 
\{ \ell_{x,\sigma}^* , \ell_{y,\tau}^* \} 
\ = \ 
\{ \ell_{x,\sigma} , \ell_{y,\tau} \} 
\ = \ 
\{ h_{x,\sigma}^* , \ell_{y,\tau}^* \} 
\nonumber \\ 
\ = \ & 
\{ h_{x,\sigma} , \ell_{y,\tau} \} 
\ = \
\{ h_{x,\sigma} , \ell_{y,\tau}^* \} 
\ = \ 
\{ h_{x,\sigma}^* , \ell_{y,\tau} \} 
\ = \ 0 \, , 
\\[1ex] \label{eq-III.26}
\{ h_{x,\sigma} , h_{y,\tau}^* \} \ = \ &  
\la \delta_{x,\sigma} | \, P_\HF^\perp \, \delta_{y,\tau} \ra 
\, , \ \ 
\{ \ell_{x,\sigma} , \ell_{y,\tau}^* \} \ = \ 
\ol{\la \delta_{x,\sigma} | \, P_\HF \, \delta_{y,\tau} \ra} 
\, , \ \  
\\[1ex] \label{eq-III.27}
& \qquad \quad h_{x,\sigma} \Om \ = \ \ell_{x,\sigma} \Om \ = \ 0 \, ,
\end{align}
for all $x,y \in \Lambda$ and $\sigma, \tau \in \{\udarrow\}$, and
\begin{align} \label{eq-III.28,1}
\QQ_1 \ := \ &
\sum_{x,y \in \Lambda} \sum_{\sigma, \tau \in \{\udarrow\} } 
v_{x-y} \, h_{x,\sigma}^* \, h_{y,\tau}^* \, h_{y,\tau} \, h_{x,\sigma} \, , 
\\ \label{eq-III.28,2}
\QQ_2 \ := \ &
\sum_{x,y \in \Lambda} \sum_{\sigma, \tau \in \{\udarrow\} } 
v_{x-y} \, \ell_{y,\tau}^* \, \ell_{x,\sigma}^* \, \ell_{x,\sigma} \, \ell_{y,\tau} \, ,
\\ \label{eq-III.28,3}
\QQ_3 \ := \ & 
\sum_{x,y \in \Lambda} \sum_{\sigma, \tau \in \{\udarrow\} } 
v_{x-y} \, h_{x,\sigma}^* \, \ell_{y,\tau}^* \, \ell_{y,\tau} \, h_{x,\sigma} \, , 
\\ \label{eq-III.28,4}
\QQ_4 \ := \ & 
\sum_{x,y \in \Lambda} \sum_{\sigma, \tau \in \{\udarrow\} } 
v_{x-y} \, h_{y,\tau}^* \, \ell_{y,\tau}^* \, \ell_{x,\sigma} \, h_{x,\sigma} \, , 
\\ \label{eq-III.28,5}
\QQ_5 \ := \ &
\sum_{x,y \in \Lambda} \sum_{\sigma, \tau \in \{\udarrow\} } 
v_{x-y} \, h_{x,\sigma}^* \, \ell_{y,\tau}^* \, \ell_{x,\sigma}^* \, \ell_{y,\tau} \, ,
\\ \label{eq-III.28,6}
\QQ_6 \ := \ & 
\sum_{x,y \in \Lambda} \sum_{\sigma, \tau \in \{\udarrow\} } 
v_{x-y} \, h_{y,\tau}^* \, h_{y,\tau} \, \ell_{x,\sigma} \, h_{x,\sigma} \, ,
\\ \label{eq-III.28,7}
\QQ_7 \ := \ & 
\sum_{x,y \in \Lambda} \sum_{\sigma, \tau \in \{\udarrow\} } 
v_{x-y} \, h_{x,\sigma}^* \, h_{y,\tau}^* \, \ell_{y,\tau}^* \, \ell_{x,\sigma}^* \, .
\end{align}
Finally, we introduce the one-particle density 
\begin{align} \label{eq-III.33}
\rho_\HF(x) \ := \ \sum_{\sigma \in \{\udarrow\}} \rho_\HF(x,\tau)
\, , \qquad  
\rho_\HF(x,\tau) \ := \ 
\la \delta_{x,\sigma} | \, P_\HF \, \delta_{x,\sigma} \ra
\end{align}
of the Hartree--Fock ground state at $x \in \Lambda$ 
and the number operators  
\begin{align} \label{eq-III.34}
\NN_h \ := \! 
\sum_{x \in \Lambda, \sigma \in \{\udarrow\}}  
h_{x,\sigma}^* \, h_{x,\sigma} 
\, , \quad
\NN_\ell \ := \!  
\sum_{x \in \Lambda, \sigma \in \{\udarrow\}}  
\ell_{x,\sigma}^* \, \ell_{x,\sigma} 
\, , \quad
\NN \ = \ \NN_h + \NN_\ell \, .
\end{align}
With these definitions we are in position to formulate the relative
bounds on $\QQ_3$, $\QQ_4$, $\QQ_5$, and $\QQ_6$ to demonstrate
that these terms are under control. We remark that the bounds 
formulated in Theorem~\ref{thm-III.01} below for $\kappa = 2$,
actually hold uniformly for $0 < \kappa \leq 2$.
%
\begin{theorem} \label{thm-III.01} The interaction terms 
$\QQ_3$, $\QQ_4$, $\QQ_5$, and $\QQ_6$ vanish on the vacuum sector
and obey the following quadratic
form bounds on the orthogonal complement of the vacuum sector:
\begin{align} \label{eq-III.35}
\big\| \NN^{-1/2} \, \QQ_3 \, \NN^{-1/2} \big\| \, , \;
\big\| \NN^{-1/2} \, \QQ_4 \, \NN^{-1/2} \big\| \, , \;
\big\| \NN^{-1/2} \, \QQ_5 & \, \NN^{-1/2} \big\| 
\ \leq \
2 \, \| v * \rho_\HF \|_\infty \, ,
\\[1ex] \label{eq-III.36}
\big\| (\NN + \QQ_1)^{-1/2} \, \QQ_6 \, (\NN + \QQ_1)^{-1/2} \big\|
\ \leq \ &
\| v * \rho_\HF \|_\infty \, ,
\end{align}
where 
$\| v * \rho_\HF \|_\infty = \max_{x \in \Lambda} 
\sum_{y \in \Lambda} v(x-y) \, \rho_\HF(y)$.
\begin{proof} We only need to bound the absolute value of 
diagonal matrix elements $\la \QQ_\nu \ra := \la \Psi | \, \QQ_\nu \,
\Psi \ra$ of normalized vectors $\Psi \in \fF$. We make multiple use
the Cauchy-Schwarz inequality 
$|\la A^* B \ra|^2 \leq \la A^* A \ra \, \la B^* B \ra$. Additionally
using 
\begin{align} \label{eq-III.39}
\ell_{y,\tau}^* \, \ell_{y,\tau} 
\ \leq \ 
\rho_\HF(y,\tau) \cdot \bfone_\fF
\end{align}
for the estimate of $\QQ_3$, we obtain
\begin{align} \label{eq-III.40}
\big| \la \QQ_3 \ra \big|
\ = \ & 
2 \sum_{x,y \in \Lambda} \sum_{\sigma, \tau \in \{\udarrow\} } v_{x-y} 
\la h_{x,\sigma}^* \ell_{y,\tau}^* \, \ell_{y,\tau} \, h_{x,\sigma} \ra 
\nonumber \\[1ex]
\ \leq \ & 
2 \sum_{x,y \in \Lambda} \sum_{\sigma \in \{\udarrow\} } v_{x-y} \, \rho_\HF(y) \,
\la h_{x,\sigma}^* \, h_{x,\sigma} \ra 
\nonumber \\[1ex]
\ \leq \ & 
2 \| v * \rho_\HF \|_\infty \; \la \NN_h \ra 
\ \leq \  
2 \| v * \rho_\HF \|_\infty \; \la \NN \ra \, .
\end{align}
By the Cauchy-Schwarz inequality and again \eqref{eq-III.39}, we have
\begin{align} \label{eq-III.41}
\big| \la \QQ_5 \ra \big|
\ = \ & 
\sum_{x,y \in \Lambda} \sum_{\sigma, \tau \in \{\udarrow\} } v_{x-y} 
\big| \la h_{x,\sigma}^* \, \ell_{y,\tau}^* \, 
\ell_{x,\sigma}^* \, \ell_{y,\tau} \ra \big|
\nonumber \\[1ex]
\ \leq \ & 
\sum_{x,y \in \Lambda} \sum_{\sigma, \tau \in \{\udarrow\} } v_{x-y} 
\la h_{x,\sigma}^* \, \ell_{y,\tau}^* \, \ell_{y,\tau} \, h_{x,\sigma} \ra^{1/2}
\;  \la \ell_{y,\tau}^* \, \ell_{x,\sigma} \, 
\ell_{x,\sigma}^* \, \ell_{y,\tau} \ra^{1/2}
\nonumber \\[1ex]
\ \leq \ & 
\sum_{x,y \in \Lambda} \sum_{\sigma, \tau \in \{\udarrow\} } v_{x-y} 
\sqrt{ \rho_\HF(x,\sigma) \, \rho_\HF(y,\tau) \: } \:
\la h_{x,\sigma}^* \, h_{x,\sigma} \ra^{1/2}
\, \la \ell_{y,\tau}^* \, \ell_{y,\tau} \ra^{1/2}
\nonumber \\[1ex]
\ \leq \ & 
\| v * \rho_\HF \|_\infty \; \la \NN_h \ra^{1/2} \, \la \NN_\ell \ra^{1/2} 
\ \leq \  
\| v * \rho_\HF \|_\infty \; \la \NN \ra \, .
\end{align}
Next, we observe that $\QQ_4 = \QQ_4' - \QQ_4''$, where 
\begin{align} \label{eq-III.42}
\QQ_4' \ := \ &
\sum_{x,y \in \Lambda} \sum_{\sigma, \tau \in \{\udarrow\} } v_{x-y} 
\la \delta_{x,\sigma} | \, P_\HF \, \delta_{y,\tau} \ra
\la h_{y,\tau}^* \, h_{x,\sigma} \ra \, ,
\\[1ex] \label{eq-III.43}
\QQ_4'' \ := \ &
\sum_{x,y \in \Lambda} \sum_{\sigma, \tau \in \{\udarrow\} } v_{x-y} 
\la h_{y,\tau}^* \, \ell_{x,\sigma} \, \ell_{y,\tau}^* \, h_{x,\sigma} \ra \, ,
\end{align}
and thanks to 
$|\la \delta_{x,\sigma} | \, P_\HF \, \delta_{y,\tau} \ra|^2 \leq 
\rho_\HF(x,\sigma) \rho_\HF(y,\tau)$,
these two terms obey the estimates
\begin{align} \label{eq-III.44}
\big| \la \QQ_4' \ra \big| 
\ \leq \ &
\sum_{x,y \in \Lambda} \sum_{\sigma, \tau \in \{\udarrow\} } v_{x-y} 
\sqrt{ \rho_\HF(x,\sigma) \, \rho_\HF(y,\tau) \: } \:
\la h_{y,\tau}^* \, h_{y,\tau} \ra^{1/2} \, 
\la h_{x,\sigma}^* \, h_{x,\sigma} \ra^{1/2}
\nonumber \\[1ex]
\ \leq \ & 
\| v * \rho_\HF \|_\infty \; \la \NN_h \ra 
\end{align}
and
\begin{align} \label{eq-III.45}
\big| \la \QQ_4'' \ra \big| 
\ \leq \ &
\sum_{x,y \in \Lambda} \sum_{\sigma, \tau \in \{\udarrow\} } v_{x-y} 
\la h_{y,\tau}^* \, \ell_{x,\sigma} \, \ell_{x,\sigma}^* \, h_{y,\tau} \ra^{1/2}
\; \la h_{x,\sigma}^* \, \ell_{y,\tau} \, \ell_{y,\tau}^* \, h_{x,\sigma} \ra^{1/2}
\nonumber \\[1ex]
\ \leq \ & 
\| v * \rho_\HF \|_\infty \; \la \NN_h \ra \, 
\end{align}
which yields 
$|\la \QQ_4 \ra| \leq 2 \|v * \rho_\HF\|_\infty \la \NN \ra$. 
Finally, 
\begin{align} \label{eq-III.46}
\big| \la \QQ_6 \ra \big| 
\ \leq \ &
\sum_{x,y \in \Lambda} \sum_{\sigma, \tau \in \{\udarrow\} } 
v_{x-y} \, \big| \la 
h_{y,\tau}^* \, \ell_{x,\sigma} \, h_{y,\tau} \, h_{x,\sigma} \ra \big| 
\nonumber \\[1ex]
\ \leq \ & 
\sum_{x,y \in \Lambda} \sum_{\sigma, \tau \in \{\udarrow\} } v_{x-y} 
\la h_{y,\tau}^* \, \ell_{x,\sigma} \, \ell_{x,\sigma}^* \, h_{y,\tau} \ra^{1/2}
\; \la h_{x,\sigma}^* \, h_{y,\tau}^* \, h_{y,\tau} \, h_{x,\sigma} \ra^{1/2}
\nonumber \\[1ex]
\ \leq \ & 
\| v * \rho_\HF \|_\infty^{1/2} \: \la \NN_h \ra^{1/2} 
\, \la \QQ_1 \ra^{1/2} \, .
\end{align}
\end{proof}
\end{theorem}

\paragraph*{Main Result: Lower Bound on 
$\boldsymbol{|\la \QQ_7 \ra|}$.} We come to the main result of this
paper, namely a lower bound on the absolute value of $\la \Psi | \QQ_7
\Psi \ra$, for a suitable choice of $\Psi$, which proves that $\QQ_7$
neither obeys a quadratic form bound \eqref{eq-III.35} nor
\eqref{eq-III.36} nor any other bound that is nontrivial in the limit
$D \to \infty$.

The absence of such a bound does not hold in general, and of course, a
counterexample depends on the model. The counterexample we
give is based on the Hartree--Fock ground state of the Hubbard model
at half-filling because in this case the solution is explicitly known
\cite{BachLiebSolovej1994} - we review its construction below. We
point out that this additionally illustrates that the absence of a
relative bound for $\QQ_7$ is not caused by the long-range nature of
the interaction potential - in fact, in the Hubbard model the pair
interaction is $v_{x-y} = \delta_{x,y}$, i.e., of zero range.

Before we focus on the Hubbard model, we characterize our choice of
$\Psi$ and the main term $\la \Psi | \QQ_7 \Psi \ra$ it yields in the
following theorem which, like Theorem~\ref{thm-III.01}, we formulate
only for $\kappa = 2$ - even though it actually holds true uniformly
for all $0 < \kappa \leq 2$.
%
\begin{theorem} \label{thm-III.02} 
Assume that $t \in \cB[\fh]$ and $v \in \cB[\fh \otimes \fh]$ are 
bounded uniformly in $D = \dim(\fh)$, and define by 
\begin{align} \label{eq-III.46,01}
v_\wedge \ := \ & \frac{1}{4} \, (1 - \Ex) \, v \, (1 - \Ex) 
\ \in \ \cB[\fh \otimes \fh]
\end{align}
the restriction of $v$ to the subspace 
$\fh \wedge \fh \subseteq \fh \otimes \fh$ of antisymmetric vectors.
For $\QQ_7$ as in \eqref{eq-III.17} and $\eps \in (0,\frac{1}{2}]$, 
define the normalized vector
\begin{align} \label{eq-III.47}
\Phi_\eps \ := \ &
\sqrt{1-\eps^2} \, \Om \: + \: \eps \, \|\QQ_7 \Om\|^{-1} \, \QQ_7 \Om \, .
\end{align}
Then 
\begin{align} \label{eq-III.48,01}
\la \Phi_\eps | \TT_\HF \Phi_\eps \ra
\ \leq \ &
\|t\|_\op \, \la \Phi_\eps | \NN \Phi_\eps \ra
\ \leq \
4 \, \eps^2 \, \|t\|_\op \, ,
\\[1ex] \label{eq-III.48,02}
\la \Phi_\eps | \QQ_\main \Phi_\eps \ra
\ \leq \ &
4 \, \eps^2 \, \|v\|_\op \, ,
\\[1ex] \label{eq-III.48,03}
\la \Phi_\eps | \QQ_7 \Phi_\eps \ra
\ = \ &
2 \, \eps \, \|\QQ_7 \Om\| \, ,
\end{align}
where $\TT_\HF$ is as in \eqref{eq-III.12}, and the number operator
$\NN$ is defined in \eqref{eq-III.08,01}. In particular, choosing
$\eps := \min\{\frac{1}{2} , (1+\|v\|_\op)\} > 0$, we have that
\begin{align} \label{eq-III.48,04}
\frac{ \la \Phi_\eps | \QQ_7 \Phi_\eps \ra}{
\la \Phi_\eps | (\TT_\HF + \QQ_\main + \bfone) \Phi_\eps \ra}
\ \geq \ 
\frac{\min\big\{ \tfrac{1}{2} \, , \, \sqrt{1 + \|v\|_\op } 
\big\}}{4 + \|t\|_\op} \, 
\|\QQ_7 \Om\| \, .
\end{align}
Furthermore, 
\begin{align} \label{eq-III.48,05}
\| \QQ_7 \Om \|^2 
\ = \ 
\Tr_{\fh \otimes \fh}\big[ v_\wedge \, (P_\HF^\perp \otimes P_\HF^\perp) \, 
v_\wedge \, (P_\HF \otimes P_\HF) \big] \, .
\end{align}
\begin{proof} We first recall that
\begin{align} \label{eq-III.49,01}
\QQ_7 \ = \ &
\sum_{j, k \in \cIh} \ \sum_{m, n \in \cIl} 
V_{j, k; m,n} \, h_k^* h_j^* \ell_m^* \ell_n^* \, ,
\\ \label{eq-III.49,02}
\QQ_1 \ = \ & 
\sum_{r, s, t, u \in \cIh} V_{r, s; t,u} \, h_s^* h_r^* h_t h_u \, , 
\\ \label{eq-III.49,03}
\QQ_2 \ = \ & 
\sum_{r, s, t, u \in \cIl} V_{r, s; t,u} \, \ell_t^* \ell_u^* \ell_s \ell_r \, ,
\end{align}
from \eqref{eq-III.14} and \eqref{eq-III.17}. Observe that 
$\QQ_7 \Om \perp \Om$ because $\NN \Om = 0$ and 
$\NN \QQ_7 \Om = 4 \QQ_7 \Om$ belong to different particle number
subspaces. Hence, $\Phi_\eps$ is normalized and
\begin{align} \label{eq-III.50}
0 \ \leq \ \la \Phi_\eps | \TT_\HF \Phi_\eps \ra 
\ \leq \ 
\|t\|_\op \, \la \Phi_\eps | \NN \Phi_\eps \ra 
\ = \ 
4 \, \eps^2 \, \|t\|_\op \, .
\end{align}
Furthermore, $\QQ_\main := \QQ_1 + \QQ_2$ preserves the particle number
and $\QQ_\main \Om = 0$. Thus
\begin{align} \label{eq-III.50,1}
\la \Phi_\eps | \QQ_\main \Phi_\eps \ra 
\ = \ 
\eps^2 \, \| \QQ_7 \Om \|^{-2} \, \big( 
\la \Om | \QQ_7^* \, \QQ_1 \, \QQ_7 \Om \ra \, + \,
\la \Om | \QQ_7^* \, \QQ_2 \, \QQ_7 \Om \ra \big) \, .
\end{align}
Similarly, we obtain from 
$\la \Om | \QQ_7 \Om \ra = \la \QQ_7 \Om | \QQ_7^2 \Om \ra = 0$ that
\begin{align} \label{eq-III.51}
\la \Phi_\eps | \rRe[\QQ_7] \Phi_\eps \ra 
\ = \ &
\eps \, \| \QQ_7 \Om \|^{-1} \, 
\big( \la \Om | (\QQ_7 + \QQ_7^*) \QQ_7 \Om \ra
\, + \, \la \QQ_7 \Om | (\QQ_7 + \QQ_7^*) \Om \ra \big) 
\nonumber \\[1ex]
\ = \ &
2 \eps \, \| \QQ_7 \Om \|^{-1} \, \la \Om | \QQ_7^* \QQ_7 \Om \ra 
\ = \
2 \eps \, \| \QQ_7 \Om \| \, .
\end{align}
Next, we compute $\la \Om | \QQ_7^* \, \QQ_7 \Om \ra$, 
$\la \Om | \QQ_7^* \, \QQ_1 \, \QQ_7 \Om \ra$, and 
$\la \Om | \QQ_7^* \, \QQ_2 \, \QQ_7 \Om \ra$. To this end we use 
\eqref{eq-III.07} and \eqref{eq-III.08} and obtain
\begin{align} \label{eq-III.53,01}
\la \Om | \ell_{n'} \ell_{m'} h_{j'} h_{k'} \, 
h_k^* h_j^* \ell_m^* \ell_n^* \Om \ra 
\ = \ &
\la \Om | h_{j'} h_{k'} \, h_k^* h_j^* \Om \ra \, 
\la \Om | \ell_{n'} \ell_{m'} \, \ell_m^* \ell_n^* \Om \ra \, ,
\\[1ex] \label{eq-III.53,02} 
\la \Om | h_{j'} h_{k'} \, h_k^* h_j^* \Om \ra 
\ = \ &
(\delta_{j, j'} \delta_{k, k'} - \delta_{j, k'} \delta_{k, j'}) \, ,
\\[1ex] \label{eq-III.53,03} 
\la \Om | \ell_{n'} \ell_{m'} \, \ell_m^* \ell_n^* \Om \ra 
\ = \ &
( \delta_{m, m'} \delta_{n, n'} - \delta_{m, n'} \delta_{n, m'}) \, ,
\end{align}
for all $j, j', k, k' \in \cIh$ and $m, m', n, n' \in \cIl$.
Moreover, if additionally $t, u \in \cIh$ and $r, s \in \cIl$
then
\begin{align} \label{eq-III.53,04}
h_t h_u h_k^* h_j^* \ell_m^* \ell_n^* \Om
\ = \ &
(\delta_{u,k} \delta_{t,j} - \delta_{u,j} \delta_{t,k}) 
\ell_m^* \ell_n^* \Om \, , 
\\[1ex] \label{eq-III.53,05}
\ell_s \ell_r h_k^* h_j^* \ell_m^* \ell_n^* \Om
\ = \ &
(\delta_{r,m} \delta_{s,n} - \delta_{r,n} \delta_{s,m}) 
h_k^* h_j^* \Om \, , 
\end{align}
which imply
\begin{align} \label{eq-III.53,06}
\la \Om | \ell_{n'} \ell_{m'} & h_{j'} h_{k'} \, h_s^* h_r^* h_t h_u \,
h_k^* h_j^* \ell_m^* \ell_n^* \Om \ra 
\ = \ 
\la h_r h_s h_{k'}^* h_{j'}^* \ell_{m'}^* \ell_{n'}^* \Om | 
h_t h_u h_k^* h_j^* \ell_m^* \ell_n^* \Om \ra 
\nonumber \\[1ex] 
\ = \ &
(\delta_{s,k'} \delta_{r,j'} - \delta_{s,j'} \delta_{r,k'}) 
(\delta_{u,k} \delta_{t,j} - \delta_{u,j} \delta_{t,k}) 
\la \ell_{m'}^* \ell_{n'}^* \Om | \ell_m^* \ell_n^* \Om \ra 
\\[1ex] \nonumber 
\ = \ &
(\delta_{s,k'} \delta_{r,j'} - \delta_{s,j'} \delta_{r,k'}) 
(\delta_{u,k} \delta_{t,j} - \delta_{u,j} \delta_{t,k}) 
(\delta_{m, m'} \delta_{n, n'} - \delta_{m, n'} \delta_{n, m'}) \, ,
\end{align}
and, similarly, 
\begin{align} \label{eq-III.53,07}
\la \Om | \ell_{n'} \ell_{m'} & h_{j'} h_{k'} \, \ell_t^* \ell_u^* \ell_s \ell_r 
\, h_k^* h_j^* \ell_m^* \ell_n^* \Om \ra 
\\[1ex] \nonumber 
\ = \ &
(\delta_{r,m} \delta_{s,n} - \delta_{r,n} \delta_{s,m}) 
(\delta_{t,m'} \delta_{u,n'} - \delta_{t,n'} \delta_{u,m'}) 
(\delta_{k, k'} \delta_{j, j'} - \delta_{k, j'} \delta_{j, k'}) \, .
\end{align}
Eqs.~\eqref{eq-III.53,01}-\eqref{eq-III.53,07} yield
\begin{align} \label{eq-III.52,01}
\| \QQ_7 \Om & \|^2 
\nonumber \\ 
\ = \ & 
\sum_{j, k, j', k' \in \cIh} \ \sum_{m, n, m', n' \in \cIl} 
\ol{V_{j', k'; m',n'}} \, V_{j, k; m,n} \,  
\big\la \Om \big| \, \ell_{n'} \ell_{m'} h_{j'} h_{k'} \, 
h_k^* h_j^* \ell_m^* \ell_n^* \Om \big\ra 
\nonumber \\[1ex] 
\ = \ &
2 \sum_{j, k \in \cIh} \ \sum_{m, n \in \cIl} 
V_{m,n; j,k} \, ( V_{j,k; m,n} \, - \, V_{k,j; m,n} )
\\[1ex] \nonumber 
\ = \ &
2 \, \Tr_{\fh \otimes \fh}\big[ v_\wedge \, (P_\HF^\perp \otimes P_\HF^\perp) 
\, v_\wedge \, (P_\HF \otimes P_\HF) \big] \, ,
\end{align}
where we use \eqref{eq-III.04,1}, $V_{k,j; n,m} = V_{j,k; m,n}$ and
$\ol{V_{j,k; m,n}} = V_{m,n; j,k}$, which follow from the
symmetry $\Ex \circ v = v \circ \Ex$ and the self-adjointness of $v$.
Similarly, 
\begin{align} \label{eq-III.52,02}
\la \Om | \QQ_7^* \, & \QQ_1 \, \QQ_7 \Om \ra 
\nonumber \\ 
\ = \ & 
4 \sum_{j, k, r, s \in \cIh} \ \sum_{m, n \in \cIl} 
\ol{V_{r, s ; m, n}} \, V_{r, s; j, k} \, 
\big( V_{j, k; m, n} \, - \, V_{j, k; n, m} \big)
\nonumber \\[1ex] 
\ = \ &
4 \, \Tr_{\fh \otimes \fh}\big[ v_\wedge \, (P_\HF^\perp \otimes P_\HF^\perp) 
\, v_\wedge \, (P_\HF^\perp \otimes P_\HF^\perp) \, v_\wedge \, 
(P_\HF \otimes P_\HF) \big] \, ,
\end{align}
and 
\begin{align} \label{eq-III.52,03}
\la \Om | \QQ_7^* \, & \QQ_2 \, \QQ_7 \Om \ra 
\nonumber \\ 
\ = \ & 
4 \sum_{j, k, \in \cIh} \ \sum_{m, n, t, u \in \cIl} 
\ol{V_{j, k; t, u}} \, V_{m, n; t, u} \, 
\big( V_{j, k; m, n} \, - \, V_{j, k; n, m} \big)
\nonumber \\[1ex] 
\ = \ &
4 \, \Tr_{\fh \otimes \fh}\big[ v_\wedge \, (P_\HF \otimes P_\HF) 
\, v_\wedge \, (P_\HF^\perp \otimes P_\HF^\perp) \, v_\wedge \, 
(P_\HF \otimes P_\HF) \big] \, .
\end{align}
We abbreviate the two orthogonal projections 
$P_\HF \otimes P_\HF =: \PP$ and $P_\HF^\perp \otimes P_\HF^\perp =: \PP_\perp$,
observing that $\PP + \PP_\perp \neq \bfone$. With the abbreviations,
we further introduce
\begin{align} \label{eq-III.52,04}
A_1 \ := \ \PP_\perp v_\wedge \PP_\perp \, , \quad & 
B_1 \ := \ \PP_\perp v_\wedge \PP v_\wedge \PP_\perp \ \geq \ 0 \, ,
\\[1ex] \label{eq-III.52,05}
A_2 \ := \ \PP v_\wedge \PP \, , \ \ \quad & \ 
B_2 \ := \ \PP v_\wedge \PP_\perp v_\wedge \PP \ \geq \ 0 \, .
\end{align}
Then, for $\nu = 1, 2$, we have that
\begin{align} \label{eq-III.52,05}
\la \Om | \QQ_7^* \, \QQ_\nu \, \QQ_7 \Om \ra
\ = \ &
4 \, \Tr_{\fh \otimes \fh}[ A_\nu \, B_\nu]
\ \leq \ 
4 \, \| A_\nu \|_\op \Tr_{\fh \otimes \fh}[ B_\nu]
\nonumber \\[1ex]
\ \leq \ &
2 \, \| v \|_\op \| \QQ_7 \Om \|^2 \, ,
\end{align}
and thus $\la \Om | \QQ_7^* \, \QQ_\main \, \QQ_7 \Om \ra \leq 
4 \| v \|_\op \| \QQ_7 \Om \|^2$. 
Eq.~\eqref{eq-III.48,04} finally results from putting the latter
estimate together with \eqref{eq-III.50,1}, \eqref{eq-III.51}, 
and \eqref{eq-III.50}. 
\end{proof}
\end{theorem}

We are now in position to formulate our main assertion on the absence
of uniform relative bounds on the example of the Hubbard model
at half-filling.
%
\begin{theorem}[Absence of Uniform Relative Bounds] \label{thm-III.03} 
Let $d \in \ZZ^+$, $L \in 4\ZZ^+$, $\Lambda = \ZZ_L^d$, 
$\Lambda^* = \frac{2\pi}{L}\ZZ_L^d$, and $g >0$. For the Hubbard
model at half-filling described in Section~\ref{sec-IV} below, 
it holds true that 
\begin{align} \label{eq-III.54,01}
\| \QQ_7 \Om \|^2 \ \geq \ \frac{a}{2} \, |\Lambda| \, ,
\end{align}
where
\begin{align} \label{eq-III.54,02}
a \ := \ &
\frac{1}{|\Lambda^*|} \sum_{\xi \in \Lambda^*} 
\frac{\om_\xi^2}{\om_\xi^2 + (g/2)^2 } 
\ \geq \
\frac{1}{4^d} \frac{d^2}{d^2 + g^2 } \, . 
\end{align}
\begin{proof} We first notice that the Hubbard model falls
into the category of translation invariant models specified in 
\eqref{eq-III.22}-\eqref{eq-III.34}. According to \eqref{eq-III.28,7},
in this case $\QQ_7$ takes the simple form
\begin{align} \label{eq-III.55}
\QQ_7 \ = \ &
\sum_{x \in \Lambda}
2 h_{x,\uparrow}^* h_{x,\downarrow}^* \ell_{x,\downarrow}^* \ell_{x,\uparrow}^* \, .
\end{align}
Hence
\begin{align} \label{eq-III.56}
\| \QQ_7 \Om \|^2 
\ = \ & 
4 \sum_{x \in \Lambda} 
\big\la \Om \big| \, 
\ell_{x,\uparrow} \ell_{x,\downarrow} h_{x,\downarrow} h_{x,\uparrow} \, 
h_{x,\uparrow}^* h_{x,\downarrow}^* \ell_{x,\downarrow}^* \ell_{x,\uparrow}^* 
\Om \big\ra 
\\[1ex] \nonumber 
\ = \ & 
4 \sum_{x \in \Lambda} 
\Big( \la \delta_{x,\uparrow} | P_\HF^\perp \delta_{x,\uparrow} \ra \, 
\la \delta_{x,\downarrow} | P_\HF^\perp \delta_{x,\downarrow} \ra 
\: - \: 
\big| \la \delta_{x,\uparrow} | P_\HF^\perp \delta_{x,\downarrow} \ra \big|^2 
\Big)
\\ \nonumber 
& \qquad \cdot 
\Big( \la \delta_{x,\uparrow} | P_\HF \delta_{x,\uparrow} \ra \, 
\la \delta_{x,\downarrow} | P_\HF \delta_{x,\downarrow} \ra 
\: - \: 
\big| \la \delta_{x,\uparrow} | P_\HF \delta_{x,\downarrow} \ra \big|^2 
 \Big) \, .
\end{align}
Introducing the self-adjoint $2 \times 2$ matrix $M_x(\sigma, \tau) :=
\la \delta_{x,\sigma}| P_\HF \delta_{x,\tau} \ra$, we observe that
$\Tr_{\CC^2}[M_x] = \Tr_{\fh}[ (\bfone_x \otimes \bfone_{\CC^2}) P_\HF
] = 1$, due to \eqref{eq-IV.30}. Thus, $\det[\bfone - M_x] =
\det[M_x]$ and
\begin{align} \label{eq-III.57}
\| \QQ_7 \Om \|^2 
\ = \ & 
4 \sum_{x \in \Lambda} 
\Big( \la \delta_{x,\uparrow} | P_\HF \delta_{x,\uparrow} \ra \, 
\la \delta_{x,\downarrow} | P_\HF \delta_{x,\downarrow} \ra 
\: - \: 
\big| \la \delta_{x,\uparrow} | P_\HF \delta_{x,\downarrow} \ra \big|^2 
 \Big)^2 \, .
\end{align}
Moreover, $\la \delta_{x,\uparrow} | P_\HF \delta_{x,\downarrow} \ra =
0$ according to \eqref{eq-IV.39,04}. Inserting 
$\la \delta_{x,\uparrow} | P_\HF \delta_{x,\uparrow} \ra$ and 
$\la \delta_{x,\downarrow} | P_\HF \delta_{x,\downarrow} \ra$ from
\eqref{eq-IV.39,04}, we thus arrive at
\begin{align} \label{eq-III.58}
\| \QQ_7 \Om \|^2 
\ = \ & 
\sum_{x \in \Lambda} \big( 1 \, - \, 4 \Delta^2 \big)
\ = \ 
\big( 1 \, - \, 4 \Delta^2 \big) \, |\Lambda| \, ,
\end{align}
where $0 < \Delta < \frac{1}{2}$ is the unique solution of
\eqref{eq-IV.14,2}. It remains to show that $\Delta < \frac{1}{2}$
uniformly in $L \to \infty$. To this end it is convenient to
introduce
\begin{align} \label{eq-IV.59}
\EE[ X ] \ := \ \frac{1}{|\Lambda^*|} \sum_{\xi \in \Lambda^*} X_\xi 
\, , \quad
0 \ < \ \eps \ := \ 1 \: - \: 4 \, \Delta^2 \ < \ 1
\quad \text{and} \quad 
\hatom_\xi \ := \ \frac{2 \om_\xi}{g} \, ,
\end{align}
so that \eqref{eq-IV.14,2} is equivalent to
\begin{align} \label{eq-IV.60}
1 \ = \ & 
\frac{1}{|\Lambda^*|} \sum_{\xi \in \Lambda^*} 
\sqrt{ ( 1 - \eps + \hatom_\xi^2 )^{-1} \, } 
\ = \ 
\EE\Big[ \sqrt{ ( 1 - \eps + \hatom^2 )^{-1} \, } \Big] \, .
\end{align}
The concavity of $\lambda \mapsto \sqrt{\lambda}$ and Jensen's
inequality imply that
\begin{align} \label{eq-IV.61}
0 \ = \ & 
\bigg( \EE\Big[ \sqrt{ ( 1 - \eps + \hatom^2 )^{-1} \, } \Big] \bigg)^2
\: - \: 1
\ \leq \ 
\EE\big[ ( 1 - \eps + \hatom^2 )^{-1} \big] \: - \: 1
\nonumber \\[1ex]
\ \leq \ &
\EE\bigg[ \frac{\eps - \hatom^2}{1 - \eps + \hatom^2} \bigg]
\ \leq \ 
\frac{\eps}{1 - \eps} \: - \: 
\EE\bigg[ \frac{\hatom^2}{1 + \hatom^2} \bigg] \, .
\end{align}
Solving this inequality for $\eps$, we arrive at
\begin{align} \label{eq-IV.62}
\eps \ \geq \ \frac{a}{1+a} \ \geq \ \frac{a}{2} \, , 
\quad \text{with} \quad
a \ := \  \EE\bigg[ \frac{\hatom^2}{1 + \hatom^2} \bigg] 
\ \in \ (0,1) \, ,
\end{align}
and, hence, at \eqref{eq-III.54,01}. For the derivation of
\eqref{eq-III.54,02} we observe that 
$\cos(\frac{2 \pi n_\nu}{L}) \geq \cos(\frac{\pi}{4}) \geq \frac{1}{2}$, 
for all $n_\nu + L \ZZ \in \ZZ_L$ with $|n_\nu| \leq L/8$. 
For each coordinate direction $\nu = 1, \ldots, d$, there are
at least $L/4 \in \ZZ^+$ such $n_\nu$. Therefore, 
\begin{align} \label{eq-III.63}
a \ = \ &
\frac{1}{|\Lambda^*|} \sum_{\xi \in \Lambda^*} 
\frac{\om_\xi^2}{\om_\xi^2 + (g/2)^2 } 
\ \geq \
\frac{d^2}{d^2 + g^2 } \, 
\frac{|\{ \xi \in \Lambda^*: \, |\om_\xi| \geq \tfrac{d}{2} \}|}{L^d} 
\ \geq \
\frac{1}{4^d} \frac{d^2}{d^2 + g^2 } \, . 
\end{align}
\end{proof}
\end{theorem}

\secction{Hartree--Fock Theory of the Hubbard Model at Half-Filling} 
\label{sec-IV}
%
\paragraph*{The Hubbard Model at Half-Filling.} The Hubbard model is
a simplified model for the description of interacting electrons on a a
discrete set $\Lambda$ called the lattice. The single-fermion Hilbert
space for this model is $\fh := \ell^2(\Lambda \times \{\udarrow\})$.
Note that $\fh \cong \fg \otimes \CC^2$, where 
$\fg := \ell^2(\Lambda)$ is the space of complex-valued functions
on $\Lambda$, and we frequently change between these representations
without further notice.

Here we choose the lattice $\Lambda$ to be the discrete
$d$-dimensional torus given by $\Lambda \equiv \Lambda_L := \ZZ_L^d$,
where $\ZZ_L := \ZZ / L \ZZ$ and $L \in 4\ZZ^+$ is a positive integer
multiple of $4$. The (Pontryagin) dual lattice is $\Lambda^* \equiv
\Lambda_L^* = \frac{2\pi}{L} \ZZ_L^d$.
The lattice $\Lambda$ is a metric space w.r.t.\ the natural metric
$|x-y| := \min\big\{ |\vz - L \vq|_1 : \; \vq \in \ZZ^d \big\}$, where
$\vz \in \ZZ^d$ is such that $x-y = \vz + L\ZZ$. Similarly,
$|\xi-\eta| := \min\big\{ |\vkappa - 2\pi \vq|_\infty : \; \vq \in
\ZZ^d \big\}$ defines a metric on $\Lambda^*$.

The canonical ONB with respect to coordinate space $\Lambda$ and
Fourier space $\Lambda^*$, respectively, are 
$\{ \delta_{x,\sigma} \}_{(x,\sigma) \in \Lambda \times \{\udarrow\}} \subseteq \fh$
and 
$\{ \vphi_{\xi,\sigma} \}_{(\xi,\sigma) \in \Lambda^* \times \{\udarrow\}} \subseteq \fh$,
where $\delta_{x,\sigma}, \vphi_{\xi,\sigma} \in \fh$ are given by
\begin{align} \label{eq-IV.01}
\delta_{x,\sigma}(y,\tau) \; := \; \delta_{x,y} \, \delta_{\sigma,\tau} 
\quad \text{and} \quad
\vphi_{\xi,\sigma}(y,\tau) \; := \; 
\frac{e^{-i\xi \cdot y}}{\sqrt{ |\Lambda| }} \, \delta_{\sigma,\tau} \, ,
\end{align}
for all $(y,\tau) \in \Lambda \times \{\udarrow\}$, where 
$\xi \cdot y = \xi_1 y_1 + \ldots + \xi_d y_d$, as
usual. The fermion creation and annihilations operators corresponding
to \eqref{eq-II.05} are denoted by $c_{x,\sigma}^* :=
c^*(\delta_{x,\sigma})$ and $c_{x,\sigma} := c(\delta_{x,\sigma})$,
for $(x,\sigma) \in \Lambda \times \{\udarrow\}$, and
$\hc_{\xi,\sigma}^* := c^*(\vphi_{\xi,\sigma})$ and $\hc_{\xi,\sigma}
:= c(\vphi_{\xi,\sigma})$, for $(\xi,\sigma) \in \Lambda^* \times
\{\udarrow\}$, respectively.
 
Equipped with this notation and following \eqref{eq-Ia.01}, we
introduce the Hubbard Hamiltonian by
\begin{align} \label{eq-IV.02}
\tHH \ = \ \TT \: + \: \frac{g}{2} \, \VV \, ,
\end{align}
where, comparing with \eqref{eq-III.24}, the interaction $\VV$ is the 
on-site repulsion 
\begin{align} \label{eq-IV.03}
v_{x-y} \ = \ \delta_{x,y} \ , \quad \text{i.e.,} \quad
\VV \ := \ 
2 \sum_{x \in \Lambda} c_{x,\uparrow}^* \, c_{x,\downarrow}^* \, 
c_{x,\downarrow} \, c_{x,\uparrow} \, ,
\end{align}
and the kinetic energy 
\begin{align} \label{eq-IV.04}
\TT \ := \ 
\sum_{x,y \in \Lambda} \sum_{\sigma = \udarrow} 
t_{x-y} \: c_{x,\sigma}^* \, c_{y,\sigma} 
\ = \ 
\sum_{\xi \in \Lambda^*} \sum_{\sigma = \udarrow} 
\om_\xi \: \hc_{\xi,\sigma}^* \, \hc_{\xi,\sigma} 
\end{align}
is the second quantization of the (traceless) discrete Laplacian.
That is, $T = T^* = (t_{x,y})_{x,y \in \Lambda} \in \CC^{\Lambda
  \times \Lambda}$ is the nearest-neigbour hopping matrix and
$\om = \hat{t}$ its Fourier transform, 
\begin{align} \label{eq-IV.05}
t_z := - \bfone\big( |z| = 1 \big) 
\quad \ \text{and} \quad \ 
\om_\xi \ := \ 
\sum_{z \in \Lambda} e^{-i \xi \cdot z} \, t_z
\ = \  
- \sum_{\nu =1}^d \cos(\xi_\nu) \, .
\end{align}
Before describing the Hartree--Fock theory on the example of the
Hubbard model, we discuss the special spectral properties of the
hopping matrix $T$ that allows us to determine the Hartree--Fock
minimizers explicitly.

The hopping matrix $T$ is \textit{bipartite}, i.e., the lattice
$\Lambda = A \dot{\cup} B$ is the union of two disjoint subsets 
$A, B \subseteq \Lambda$ such that $T_{x,y} =0$ whenever either 
$x,y \in A$ or $x,y \in B$. Introducing a unitary (gauge) transformation
$G \in \cU[\ell^2(\Lambda)]$ on the functions on $\Lambda$
by 
\begin{align} \label{eq-IV.05,1}
[G \psi](x) \ := \ (-1)^x \, \psi(x) \, , 
\qquad \text{where} \quad
(-1)^x \ := \ \bfone_A(x) \: - \: \bfone_B(x) \, ,
\end{align}
it is easy to check that $G$ is an involution and that $T$ 
transforms under conjugation with $G$ as
\begin{align} \label{eq-IV.06}
G \, T \, G \ = \ - T \, .
\end{align}
This implies that the eigenvalues of $T$ come in pairs of opposite
sign and that the projections onto its negative and
positive eigenvalues, respectively, have the same dimension.

In the present case $\Lambda \equiv \Lambda_L := \ZZ_L^d$, the
subsets $A$ and $B$ are the even and odd sites, respectively,
forming a chessboard structure on $\Lambda$. More specifically,
$x = (x_1, \ldots, x_d) \in \Lambda$ belongs to $A$ or $B$ if 
$x_1 + \ldots + x_d$ is even or odd, respectively, and $G$
acts on wave functions at $x$ by multiplication with 
\begin{align} \label{eq-IV.06,01}
(-1)^x \ = \ (-1)^{x_1} \cdots (-1)^{x_d} \, .
\end{align}

In Fourier representation, $G$ acts as a translation of momenta
by $\pi := (\pi, \ldots, \pi) = -\pi \in \Lambda^*$, i.e., 
$G \vphi_{\xi,\sigma} = \vphi_{\xi+\pi,\sigma}$, which is consistent with
$\om_{\xi-\pi} = -\om_{\xi}$ when conjugating $T$ with $G$. 
Note that the translation $\xi \mapsto \xi + \pi$ is a bijection 
$\Lambda^* \to \Lambda^*$ without any fixed point. We collect the 
momenta corresponding to strictly positive eigenvalues and to strictly
negative eigenvalues, respectively, in
\begin{align} \label{eq-IV.07}
\tLambda^*_+ \ := \ 
\big\{ \xi \in \Lambda \: \big| \ \om_\xi > 0 \big\}
\qquad \text{and} \qquad 
\tLambda^*_- \ := \ 
\big\{ \xi \in \Lambda \: \big| \ \om_\xi < 0 \big\} \, ,
\end{align}
and observe that, due to $\om_{\xi-\pi} = -\om_{\xi}$, the map
$\xi \mapsto \xi + \pi$ is an involution 
$\Lambda^*_+ \to \Lambda^*_-$. Since the bijection 
$\Lambda^* \ni \xi \mapsto \xi + \pi \in \Lambda^*$ leaves 
$\tLambda_0^* := \{ \xi \in \Lambda | \ \om_\xi = 0 \} = \tLambda_0^* + \pi$
invariant, but has no fixed point, we can find a disjoint partition
$\tLambda_0^* = \tLambda_{0,+}^* \dot{\cup} \tLambda_{0,-}^*$ such that
$\xi \mapsto \xi + \pi$ is an involution 
$\tLambda_{0,-}^* \to \tLambda_{0,+}^*$. It follows that
\begin{align} \label{eq-IV.08}
\Lambda^*_+ \ := \ \tLambda^*_+ \cup \tLambda_{0,+}^* 
\qquad \text{and} \qquad 
\Lambda^*_- \ := \ \tLambda^*_- \cup \tLambda_{0,-}^* 
\end{align}
form a disjoint partition of 
$\Lambda = \Lambda^*_+ \dot{\cup} \Lambda^*_-$ such that 
$\xi \mapsto \xi + \pi$ is a bijection from $\Lambda^*_+$ to $\Lambda^*_-$
(and therefore also from $\Lambda^*_-$ to $\Lambda^*_+$). 

We are now in position to formulate the Hartree--Fock theory for the
Hubbard model \cite{BachLiebSolovej1994}. According to Lieb's
variational principle \cite{Lieb1980a, Bach1992}, the Hartree--Fock
energy of the Hubbard model for $N$ electrons is given by
\begin{align} \label{eq-IV.09}
E_\HF(N) \ := \
\inf\Big\{ \cE_\HF(\gamma) \; \Big| \ \gamma \in \cL^1(\fh) \, , \ 
0 \leq \gamma \leq \bfone \, , \ \Tr(\gamma) = N \Big\} \, ,
\end{align}
where the Hartree--Fock functional $\cE_\HF$ is defined as 
\begin{align} \label{eq-IV.10}
\cE_\HF(\gamma) \ := \ &
\Tr_\fh\big[ (T \otimes \bfone) \, \gamma \big] 
\: + \: 
g \sum_{x \in \Lambda} \sum_{\sigma, \tau = \udarrow}
\big\{ \la \delta_{x,\sigma} | \gamma \delta_{x,\sigma} \ra
\la \delta_{x,\tau} | \gamma \delta_{x,\tau} \ra
\: - \: 
|\la \delta_{x,\sigma} | \gamma \delta_{x,\tau} \ra|^2 \big\}
\nonumber \\[1ex]
\ = \ &
\Tr_\fh\big[ (T \otimes \bfone) \, \gamma \big] 
\: + \: 
g \sum_{x \in \Lambda} 
\big\{ \big[\Tr_{\CC^2}(\gamma_x)]^2 \: - \: \Tr_{\CC^2}\big( \gamma_x^2 \big) 
\big\} \, ,
\end{align}
and $\gamma_x \in \CC^{2 \times 2}$ is given as
$\gamma_x(\sigma, \tau) := \la \delta_{x,\sigma} | \gamma \delta_{x,\tau}\ra$. 

We recall that any self-adjoint matrix $A = A^* \in \CC^{2 \times 2}$
can be written as
\begin{align} \label{eq-IV.11}
A \ = \ \frac{1}{2} \Big( \rho(A) \, \bfone_{\CC^2} 
\: + \: \vv(A) \cdot \vsigma \Big) \, ,
\end{align}
where $\rho(A) := \Tr_{\CC^2}(A)$, $\vv(A) := \Tr_{\CC^2}(\vsigma\, A)$, 
and $\vsigma = (\sigma^{(1)}, \sigma^{(2)}, \sigma^{(3)})$ are the
(traceless) Pauli matrices 
$\sigma^{(1)} = \big( \begin{smallmatrix} 0 & 1 \\ 
1 & 0 \end{smallmatrix} \big)$, 
$\sigma^{(2)} = \big( \begin{smallmatrix} 0 & i \\ 
-i & 0 \end{smallmatrix} \big)$,
and $\sigma^{(3)} := \big( \begin{smallmatrix} 1 & 0 \\ 
0 & -1 \end{smallmatrix} \big)$. Since 
$\Tr(A^2) = \frac{1}{2} \rho(A) + \frac{1}{2} [\vv(A)]^2$, 
it follows that 
\begin{align} \label{eq-IV.12}
\big[ \Tr_{\CC^2}(\gamma_x) \big]^2 \: - \: \Tr_{\CC^2}\big( \gamma_x^2 \big) 
\ = \ &
\frac{1}{2} \rho_x^2 \, - \, \frac{1}{2} |\vv_x|^2 \, ,
\end{align}
where 
\begin{align} \label{eq-IV.13}
\rho_x \ := \ 
\rho(\gamma_x) 
\ = \ 
\Tr_\fh\big[ (\bfone_x \otimes \bfone) \gamma \big] 
\quad \text{and} \quad 
\vv_x \ := \ 
\vv(\gamma_x) 
\ = \ 
\Tr_\fh\big[ (\bfone_x \otimes \vsigma) \gamma \big] \, .
\end{align}
Moreover, with these definitions, 
$0 \leq \gamma_x \leq \bfone_{\CC^2}$ is equivalent to 
$0 \leq |\vv_x| \leq \rho_x \leq 2$ and we obtain
\begin{align} \label{eq-IV.14}
\cE_\HF(\gamma) \ = \ 
\Tr_\fh\big[ (T \otimes \bfone) \, \gamma \big] 
\: + \: 
\frac{g}{2} \sum_{x \in \Lambda} 
\big\{ \rho_x^2 \: - \:  |\vv_x|^2 \big\} \, .
\end{align}
Next, we characterize the Hartree--Fock energy $E_\HF(|\Lambda|)$
and Hartree--Fock ground states, i.e., 1-RDM $\gamma$ of
particle number $\Tr(\gamma) = |\Lambda|$, for which
the Hartree--Fock energy is attained, 
$\cE_\HF(\gamma) = E_\HF(|\Lambda|)$, see \cite{BachLiebSolovej1994}.
%
\begin{theorem} \label{thm-IV.01} Let $g > 0$.
\begin{itemize} 
\item[(i)] The Hartree--Fock energy per unit volume is given by
\begin{align} \label{eq-IV.14,1}
\frac{E_\HF(|\Lambda|)}{|\Lambda|}  
\ = \ & 
\frac{g}{2} \: + \: g \, \Delta^2 \: - \:  
\frac{1}{|\Lambda^*|} \sum_{\xi \in \Lambda^*} 
\sqrt{ \om_\xi^2 \: + \: g^2 \, \Delta^2 \: } \, ,
\end{align}
where $\Delta \in (0,\frac{1}{2})$ is the unique solution of 
\begin{align} \label{eq-IV.14,2}
2 \ = \ \frac{1}{|\Lambda^*|} \sum_{\xi \in \Lambda^*} 
g \, \big( \om_\xi^2 + g^2 \, \Delta^2 \big)^{-1/2} \, .
\end{align}

\item[(ii)] A reduced one-particle density matrix $\gamma \in
  \cL^1(\fh)$, $0 \leq \gamma \leq \bfone$, of particle number
  $\Tr(\gamma) = |\Lambda|$ is a Hartree-Fock ground state if, and
  only if there exists a vector $\ve \in \RR^3$ of unit length $|\ve|
  = 1$ such that
\begin{align} \label{eq-IV.14,3}
\gamma \ = \ & 
\bfone\Big[ T \otimes \bfone \: - \: 
g \Delta \, G \otimes (\ve \cdot \vsigma) \ \leq \ 0 \Big] \, .
\end{align}
\end{itemize}
\end{theorem}
%
We observe that, for any $\ve \in \RR^3$ of unit length there is a
unitary rotation $R_{\ve} \in \cU(\CC^2)$ in spin space such that
$R_{\ve} \big( \ve \cdot \vsigma \big) R_{\ve}^* = -\ve_3 \cdot
\vsigma = - \sigma^{(3)}$. Hence, for $r >0$,
\begin{align} \label{eq-IV.34}
H_r \ := \ 
T \otimes \bfone \: + \: r \, G \otimes \sigma^{(3)} 
\ = \ 
(\bfone \otimes R_{\ve}) \, [ T \otimes \bfone \, - \,
r \, G \otimes (\ve \cdot \vsigma) ] \, (\bfone \otimes R_{\ve})^* \, ,
\end{align}
and we may henceforth assume w.l.o.g.\ that $\ve = - \ve_3 =
(0,0,-1)^t$. We observe that $H_r$
leaves the two-dimensional subspaces
$\fh(\xi,\sigma) := \CC \vphi_{\xi,\sigma} \oplus \CC
\vphi_{\xi+\pi,\sigma}$ invariant, for each $(\xi,\sigma) \in
\Lambda_+^* \times \{\udarrow\}$ and all $r >0$. More specifically,
\begin{align} \label{eq-IV.35}
H_r \ = \bigoplus_{(\xi,\sigma) \in \Lambda_+^* \times \{\udarrow\}} 
\! H_r(\xi,\sigma)
\, , \quad \text{with} \qquad 
H_r(\xi,\sigma) \ = \
\begin{pmatrix} \om_\xi & \sigma r \\ 
\sigma r & - \om_\xi \end{pmatrix} 
\end{align}
w.r.t.\ the ONB $\{ \vphi_{\xi,\sigma}, \vphi_{\xi+\pi,\sigma} \}
\subseteq \fh(\xi,\sigma)$, where here and henceforth we identify
$\uparrow \; \equiv +1$ and $\downarrow \; \equiv -1$. Fixing $(\xi,\sigma)
\in \Lambda_+^* \times \{\udarrow\}$, an ONB $\{
\psi_{\xi,\sigma,+}^{(r)}, \psi_{\xi,\sigma,-}^{(r)} \} \subseteq
\fh(\xi,\sigma)$ of eigenvectors of $H_r(\xi,\sigma)$ with
corresponding eigenvalues $\pm \lambda_\xi^{(r)}$ is given by
\begin{align} \label{eq-IV.36}
\psi_{\xi,\sigma,\kappa}^{(r)} 
\ = \ 
\frac{\sigma}{\sqrt{2}} \, a_{\xi,\kappa}^{(r)} \: & \vphi_{\xi,\sigma} 
\; + \;
\frac{\kappa}{\sqrt{2}} \, a_{\xi,-\kappa}^{(r)} \: \vphi_{\xi+\pi,\sigma} \, , 
\\[1ex] \label{eq-IV.37}
a_{\xi,\kappa}^{(r)}
\ := \ &
\sqrt{ 1 + \kappa \, 
\frac{\om_\xi}{\sqrt{\smash[b]{\om_\xi^2} + r^2}} \, } \, ,
\\[1ex] \label{eq-IV.38}
\lambda_\xi^{(r)} 
\ = \ &
\sqrt{\smash[b]{\om_\xi^2} + r^2 \, } \, ,
\end{align}
for $(\xi,\sigma,\kappa) \in \Lambda_+^* \times \{\udarrow\} \times
\{\pm\}$ with $\{\pm\} := \{-1, 1\}$. We frequently omit to display
the dependence on $r >0$ and simply write $\psi_{\xi,\sigma,\kappa}
\equiv \psi_{\xi,\sigma,\kappa}^{(r)}$, $a_{\xi,\kappa} \equiv
a_{\xi,\kappa}^{(r)}$, and $\lambda_{\xi,\kappa} \equiv
\lambda_{\xi,\kappa}^{(r)}$. 

The projection onto the Hartree--Fock ground state corresponding to
$\ve = (0,0,-1)^t$ is hence given as $P_\HF = \gamma^{(g \Delta)}$ where
$\gamma^{(r)} := \bfone[ H_r \leq 0 ]$, for $r>0$. Note that 
$\gamma^{(r)} = \bfone[ H_r < 0 ]$, because all eigenvalues of $H_r$ are
nonvanishing. Moreover, we have the explicit representation
\begin{align} \label{eq-IV.39,01}
\gamma^{(r)} 
\ = \
\sum_{(\xi,\sigma) \in \Lambda_+^* \times \{\udarrow\}} 
| \psi_{\xi,\sigma,-} \ra\la \psi_{\xi,\sigma,-} | \, . 
\end{align}
Since
\begin{align} \label{eq-IV.39,02}
\la \delta_{x,\tau} | \psi_{\xi,\sigma,-} \ra
\ = \ 
\frac{\delta_{\sigma,\tau}}{\sqrt{2}} \, 
\frac{e^{-i\xi \cdot x}}{\sqrt{|\Lambda^*|}} \, 
\big[ \sigma \, a_{\xi,-} \: - \: (-1)^x \, a_{\xi,+} \big] \, , 
\end{align}
a simple computation yields
\begin{align} \label{eq-IV.39,03}
\gamma_x^{(r)}(\eta,\tau) 
\ := \
\la \delta_{x,\eta} | \gamma^{(r)}  \delta_{x,\tau} \ra
\ = \
\frac{\delta_{\sigma,\tau}}{2} \bigg[ 1 \: - \: 
\frac{\tau \, (-1)^x \, r}{|\Lambda^*|} 
\sum_{\xi \in \Lambda^*} \big( \om_\xi^2 + r^2 \big)^{-1/2} \bigg] \, . 
\end{align}
Especially for $r = g \Delta$, the self-consistent equation 
\eqref{eq-IV.14,2} implies that
\begin{align} \label{eq-IV.39,04}
\la \delta_{x,\eta} | P_\HF \delta_{x,\tau} \ra
\ = \
\delta_{\sigma,\tau} \Big( \frac{1}{2}  \: - \: 
\tau \, (-1)^x \, \Delta \Big) \, . 
\end{align}

\newpage 

\secction{APPENDIX: Proof of Theorem~\ref{thm-IV.01}}
\label{sec-A}
%
We follow \cite{BachLiebSolovej1994}. Let 
\begin{align} \label{eq-IV.14,4}
\cF_L(\eta) \ := \ & 
g \, \eta \: - \: \frac{1}{|\Lambda^*|} \sum_{\xi \in \Lambda^*} 
\sqrt{ \om_\xi^2 + g^2 \, \eta \: } \, .
\end{align}
We first show that 
$E_\HF(|\Lambda|) |\Lambda|^{-1} 
\geq \frac{g}{2} + \min_{0 \leq \eta \leq 4} \{ \cF_L(\eta) \}$.
To this end we observe that $|\Lambda|^2 = \big( \sum_{x \in \Lambda}
\rho_x \big)^2 \leq |\Lambda| \sum_{x \in \Lambda} \rho_x^2$, by the
Cauchy-Schwarz inequality. Hence, 
\begin{align} \label{eq-IV.15}
\cE_\HF(\gamma) 
\: - \: \frac{g}{2} |\Lambda| 
\ \geq \ 
\cE_\HF'(\gamma) 
\ := \ 
\Tr_\fh\big[ (T \otimes \bfone) \, \gamma \big] 
\: - \: 
\frac{g}{2} \sum_{x \in \Lambda} |\vv_x|^2 \, .
\end{align}
Note that $\cE_\HF(\gamma) - \frac{g}{2}|\Lambda| = \cE_\HF'(\gamma)$ if, 
and only if, $\rho_x =1$, for all $x \in \Lambda$. 

A trivial, but important, observation is that 
$\vv_x \geq 2 \vv_x \cdot \vw_x - \vw_x^2 \geq 0$, for any 
$\vw_x \in \RR^3$, with strict inequality unless $\vw_x = \vv_x$.
Taking $|\vv_x| \leq 2$ into account, this leads to 
$|\vv_x|^2 = \max_{|\vw_x| \leq 2} \big\{ 2 \vv_x \cdot \vw_x -
|\vw_x|^2 \big\}$ and in turn to
\begin{align} \label{eq-IV.16}
\cE_\HF'(\gamma) 
\ \geq \ &
\min_{\uw}\bigg\{ \Tr_\fh\big[ (T \otimes \bfone) \, \gamma \big] 
\: - \: \sum_{x \in \Lambda} g \, \vw_x \cdot \vv_x
\: + \: 
\frac{g}{2} \sum_{x \in \Lambda} |\vw_x|^2 \bigg\}
\nonumber \\[1ex] 
\ = \ &
\min_{\uw}\bigg\{ \Tr_\fh\bigg[ \bigg( T \otimes \bfone
\: - \: \sum_{x \in \Lambda} g \, E_{x,x} \otimes \vw_x \cdot \vsigma \bigg)
\, \gamma \bigg] 
\: + \: 
\frac{g}{2} \sum_{x \in \Lambda} |\vw_x|^2 \bigg\} \, ,
\end{align}
where $E_{x,y} \in \cB[\ell^2(\Lambda)]$ is the matrix unit
corresponding to $(x, y) \in \Lambda^2$, and $\min_{\uw}$ denotes the
minimum over 
$\uw = (\vw_x)_{x \in \Lambda} \in \ol{B_{\RR^3}(0,2)}^{|\Lambda|}$.  
Inserting this into \eqref{eq-IV.09}, we obtain the lower bound
\begin{align} \label{eq-IV.16,1}
E_\HF(|\Lambda|)  & \: - \: \frac{g}{2} |\Lambda| 
\nonumber \\ 
\ \geq \ &
\min_{\uw}\bigg\{ \inf_{0 \leq \gamma \leq \bfone}
\bigg(
\Tr_\fh\bigg[ \bigg( T \otimes \bfone
\: - \: \sum_{x \in \Lambda} g \, E_{x,x} \otimes \vw_x \cdot \vsigma \bigg)
\, \gamma \bigg] 
\: + \: 
\frac{g}{2} \sum_{x \in \Lambda} |\vw_x|^2 \bigg) \bigg\} 
\nonumber \\[1ex] 
\ = \ &
\min_{\uw}\bigg\{ \Tr_\fh\bigg[ \bigg( T \otimes \bfone \: - \: 
\sum_{x \in \Lambda} g \, E_{x,x} \otimes \vw_x \cdot \vsigma \bigg)_- 
\: \bigg] 
\: + \: 
\frac{g}{2} \sum_{x \in \Lambda} |\vw_x|^2 \bigg\} \, ,
\end{align}
where $(\lambda)_- := \min\{\lambda, 0\} = - \frac{1}{2} |\lambda| +
\frac{1}{2} \lambda = - \frac{1}{2} \sqrt{ \lambda^2 \, } +
\frac{1}{2} \lambda$ denotes the negative part of a real number
$\lambda$.  Since both $T \in \cB[\ell^2(\Lambda)]$ and
$\sigma^{(\nu)} \in \CC^{2 \times 2}$ are traceless, so is $ T \otimes
\bfone - \sum_{x \in \Lambda} g E_{x,x} \otimes \vw_x \cdot \vsigma
\in \cB[\fh]$ and hence
\begin{align} \label{eq-IV.17}
E_\HF(|\Lambda|)  & \: - \: \frac{g}{2} |\Lambda| 
\ \geq \ 
\frac{1}{2} \; \min_{\uw}\Big\{ 
\Tr_\fh\Big( - \sqrt{ A(\uw) \: } \Big)
\: + \: g \sum_{x \in \Lambda} |\vw_x|^2 \bigg\} \, ,
\end{align}
where
\begin{align} \label{eq-IV.18}
A(\uw) \ := \ &
T^2 \otimes \bfone \: + \: 
\sum_{x \in \Lambda} g^2 \, |\vw_x|^2 \, E_{x,x} \otimes \bfone
\: - \: 
g \, \bigg\{ T \otimes \bfone \; , \; 
\sum_{x \in \Lambda} E_{x,x} \otimes \vw_x \cdot \vsigma \bigg\}
\\[1ex] \nonumber 
\ = \ &
T^2 \otimes \bfone \: + \: 
\sum_{x \in \Lambda} g^2 \, |\vw_x|^2 \, E_{x,x} \otimes \bfone
\: - \: 
\sum_{x,y \in \Lambda} g \, t_{x-y} \, E_{x,y} \otimes 
( \vw_x + \vw_y) \cdot \vsigma \, ,
\end{align}
with $\{ A, B \} := AB + BA$ denoting the anticommutator
of two operators $A$ and $B$. Since $G E_{y,y} G = E_{y,y}$
and $G T G = -T$, we have that  
\begin{align} \label{eq-IV.19}
G \, A(\uw) \, G \ = \ A(-\uw) \, .
\end{align}
Furthermore, the strict convexity of 
$\RR_0^+ \ni \lambda \mapsto -\sqrt{\lambda} \in \RR$ implies
the strict convexity of 
\begin{align} \label{eq-IV.20}
A \ \mapsto \ \Tr\big[ - \sqrt{ A \, } \big] \, , 
\end{align}
as a map on self-adjoint positive operators. Eqs.~\eqref{eq-IV.19}
and \eqref{eq-IV.20} imply that
\begin{align} \label{eq-IV.21}
\Tr_\fh\Big[ -  & \sqrt{ A(\uw) \: } \Big]
\ = \
\frac{1}{2} \Tr_\fh\Big[ - \sqrt{ A(\uw) \: } \Big] 
\: + \: \frac{1}{2} \Tr_\fh\Big[ - \sqrt{ A(-\uw) \: } \Big] 
\nonumber \\[1ex]
\ \geq \ &
\Tr_\fh\bigg[ - \sqrt{ \tfrac{1}{2}A(\uw) \, + \, \tfrac{1}{2}A(-\uw) 
\,  } \bigg] 
\ = \ 
- \Tr_\fh\bigg[ \Big( T^2 \: + \: 
\sum_{x \in \Lambda} g^2 \, |\vw_x|^2 \, E_{x,x} \Big)^{1/2} \otimes \bfone 
\bigg] 
\nonumber \\[1ex]
\ = \ &
- 2 \, \Tr_\fg\bigg[ \Big( T^2 \: + \: 
\sum_{x \in \Lambda} g^2 \, |\vw_x|^2 \, E_{x,x} \Big)^{1/2} 
\bigg] \, ,
\end{align}
with strict inequality unless $t_{x-y} (\vw_x + \vw_y) = 0$, for all
$x,y \in \Lambda$. Here we use that $\fh = \fg \otimes \CC^2$, where
$\fg := \ell^2(\Lambda)$ denotes the space of complex-valued functions
on $\Lambda$.

Next, we introduce by $\tau^z \in \cU[\fg]$ the
translation of wave functions by $z \in \Lambda$. That is,
$[\tau^z \psi](x) := \psi(x-z)$, for all $x \in \Lambda$, and
$(\tau^z)^* = \tau^{-z}$. Then $\tau^z T \tau^{-z} = T$ and
$\tau^z E_{x,y} \tau^{-z} = E_{x+z,y+z}$. Again the strict convexity 
\eqref{eq-IV.20} implies that
\begin{align} \label{eq-IV.22}
- \Tr_\fg\bigg[ & \Big( T^2 \: + \: 
\sum_{x \in \Lambda} g^2 \, |\vw_x|^2 \, E_{x,x} \Big)^{1/2} \bigg] 
\nonumber \\ 
\ = \ &
- \frac{1}{|\Lambda|} \sum_{z \in \Lambda}
\Tr_\fg\bigg[ \tau^z \Big( T^2 \: + \: 
\sum_{x \in \Lambda} g^2 \, |\vw_x|^2 \, E_{x,x} \Big)^{1/2} 
\tau^{-z} \bigg] 
\nonumber \\[1ex]
\ \geq \ &
- \Tr_\fg\bigg[ \Big( T^2 \: + \: 
\frac{1}{|\Lambda|} 
\sum_{x,z \in \Lambda} g^2 \, |\vw_x|^2 \, E_{x-z,x-z} 
\Big)^{1/2} \bigg] 
\\[1ex] \nonumber 
\ = \ &
- \Tr_\fg\Big[ 
\sqrt{ T^2 \: + \: g^2 \, \la \vw^2 \ra \: } \Big] 
\ = \ 
- \sum_{\xi \in \Lambda^*} 
\sqrt{ \om_\xi^2 \: + \: g^2 \, \la \vw^2 \ra \: } \, ,
\end{align}
where $\la \vw^2 \ra := \frac{1}{|\Lambda|} \sum_{x \in \Lambda}
|\vw_x|^2 \in [0,4]$, with strict inequality unless $|\vw_x|$ is
independent of $x \in \Lambda$. Inserting this and \eqref{eq-IV.21}
into \eqref{eq-IV.17}, we arrive at $E_\HF(|\Lambda|) |\Lambda|^{-1} 
\geq \frac{g}{2} + \min_{0 \leq \eta \leq 4} \{ \cF_L(\eta) \}$
with $\cF_L: \RR_0^+ \to \RR$ as defined in \eqref{eq-IV.14,4},
\begin{align} \label{eq-IV.23}
\cF_L(\eta) 
\ := \ 
g \, \eta \: - \: \frac{1}{|\Lambda^*|} 
\sum_{\xi \in \Lambda^*} \big( \om_\xi^2 + g^2 \, \eta \big)^{1/2} \, .
\end{align}
Note that $\cF_L \in C^\infty(\RR^+;\RR)$ with
\begin{align} \label{eq-IV.24}
\cF_L'(\eta) 
\ = \ 
g \: - \: \frac{g^2}{2} \, \frac{1}{|\Lambda^*|} 
\sum_{\xi \in \Lambda^*} \big( \om_\xi^2 + g^2 \, \eta \big)^{-1/2} 
\end{align}
and $\cF_L''(\eta) > 0$, for $\eta >0$. Since $L \in 4\ZZ^+$, we have that
$\frac{1}{2}\pi = (\frac{1}{2}\pi, \ldots, \frac{1}{2}\pi) \in \Lambda^*$
with $\om_{\frac{1}{2}\pi} = \sum_{\nu=1}^d \cos(\pi/2) = 0$. This implies
that $\lim_{\eta \to 0} \cF_L'(\eta) \ = \ -\infty$. 
Furthermore, $\cF_L'(\eta) > g( 1 - \frac{1}{2\sqrt{\eta}}) \geq 0$,
for any $\eta \geq \frac{1}{4}$. It follows that
the minimum of $\cF_L$ is attained for the unique solution
$0 < \Delta^2 < \frac{1}{4}$ of \eqref{eq-IV.14,2} and that
\begin{align} \label{eq-IV.25}
\frac{E_\HF(|\Lambda|)}{|\Lambda|} 
\ \geq \ 
\frac{g}{2} \: + \: \cF_L(\Delta^2) \, .
\end{align}
Next we show that this lower bound is attained precisely by the 
projections defined in \eqref{eq-IV.14,3}. To this end we introduce
\begin{align} \label{eq-IV.26}
H(\va) \ := \ 
T \otimes \bfone \: + \: G \otimes (\va \cdot \vsigma) 
\qquad \text{and} \qquad
\gamma(\va) \ := \ \bfone\big[ H(\va) \leq 0 \big] \, ,
\end{align}
for any $\va \in \RR^3 \setminus \{0\}$. Note that zero is not an
eigenvalue of $H(\va)$ because $H(\va)^2 = (T^2 + |\va|^2) \otimes
\bfone \geq |\va|^2 > 0$. Therefore, $\gamma(\va) = \bfone[ H(\va) < 0
]$ is the projection onto the negative eigenvalues of $H(\va)$
independent of its functional form at zero, and 
\begin{align} \label{eq-IV.27}
\gamma(\va) \ = \ \frac{1}{2} \bfone - F\big[ H(\va) \big]
\end{align}
where $F \in C^\infty(\RR;\RR)$ is an odd function 
$F[-\lambda] = -F[\lambda]$ with $F \equiv 1$ on 
$( \frac{1}{2}|\va|, \infty )$. 

If $\vb \in \RR^3$ is a unit vector
perpendicular to $\va$ then $\vb \cdot \vsigma \in \cU[\CC^2]$ is a
unitary involution and $(\vb \cdot \vsigma)(\va \cdot \vsigma)(\vb
\cdot \vsigma) = -\va \cdot \vsigma$. Since furthermore $G T
G = -T$, this implies that
\begin{align} \label{eq-IV.28}
U_\vb \, H(\va) \, U_\vb \ = \ - H(\va) 
\, , \qquad 
U_\vb \ := \ U_\vb^* \ := \ 
G \otimes (\vb \cdot \vsigma) \ \in \ \cU[\fh] \, ,
\end{align}
and further 
\begin{align} \label{eq-IV.29}
\Tr_\fh\big\{ (\bfone_x \otimes \bfone) \, F[ H(\va) ] \big\}
\ = \ &
\Tr_\fh\big\{ (\bfone_x \otimes \bfone) \, 
U_\vb \, F[ H(\va) ] \, U_\vb \big\}
\ = \ 
\Tr_\fh\big\{ (\bfone_x \otimes \bfone) \, F[ -H(\va) ] \big\}
\nonumber \\[1ex]
\ = \ &
- \Tr_\fh\big\{ (\bfone_x \otimes \bfone) \, F[ H(\va) ] \big\}
\ = \ 0 \, ,
\end{align}
using that 
$\bfone_x \otimes \bfone$ and $U_\vb$ commute. Inserting this into
\eqref{eq-IV.27}, we obtain
\begin{align} \label{eq-IV.30}
\Tr_\fh[ (\bfone_x \otimes \bfone) \gamma(\va) ]
\ = \ 
\frac{1}{2} \, \Tr_\fh[ \bfone_x \otimes \bfone ]
\ = \ 1 \, , 
\end{align}
for any $x \in \Lambda$.

Next suppose that $x, y \in \Lambda$, set $z = x-y$, and use 
the unitary 
$V_z := G^{|z|} \tau^z \otimes \bfone = \tau^z G^{|z|} \otimes \bfone$
and that  
\begin{align} \label{eq-IV.31}
V_z \, F[ H(\va) ] \, V_z^*
\ = \ &
F\big[ G^{|z|} T G^{|z|} \otimes \bfone 
\: + \: \tau^z G \tau^{-z} \otimes (\va \cdot \vsigma) \big] 
\\[1ex] \nonumber 
\ = \ &
F\big[ (-1)^z T \otimes \bfone 
\: - \: (-1)^z G \otimes (\va \cdot \vsigma) \big] 
\ = \ 
F[ (-1)^z H(\va) ] 
\ = \ 
(-1)^z F[ H(\va) ] \, .
\end{align}
Since the Pauli matrices are traceless, we 
obtain
\begin{align} \label{eq-IV.32}
\Tr_\fh[ (\bfone_x & \otimes \vsigma) \gamma(\va) ]
\ = \
\Tr_\fh\big\{ (\bfone_x \otimes \vsigma)
\, F[ H(\va) ] \big\} 
\ = \
\Tr_\fh\big\{ (\bfone_y \otimes \vsigma) \, 
V_z \, F[ H(\va) ] \, V_z^* \big\} 
\nonumber \\[1ex] 
\ = \ &
- \Tr_\fh\big\{ (\bfone_y \otimes \vsigma) \, F[ H(\va) ] \big\} 
\ = \
(-1)^{y-x} \Tr_\fh[ (\bfone_y \otimes \vsigma) \gamma(\va) ] \, .
\end{align}
It follows that Inequalities~\eqref{eq-IV.15} and \eqref{eq-IV.21}
actually become \textit{equalities} when we insert $\gamma(\va) :=
\bfone\big[ T \otimes \bfone - G \otimes (\va \cdot \vsigma) \big]$
and $\vw := \vv[\gamma(\va)]$. Choosing $\va := g\Delta \ve$, this
implies that
\begin{align} \label{eq-IV.33}
\frac{E_\HF(|\Lambda|)}{|\Lambda|} 
\ \leq \ 
\frac{\cE_\HF\big( \gamma(g\Delta \ve) \big)}{|\Lambda|} 
\ = \ 
\frac{g}{2} \: + \: \cF_L(\Delta^2) 
\end{align}
and the asserted characterization (ii) of the Hartree--Fock ground states.

\newpage

\paragraph*{Acknowledgement:} We thank K.~Merz for many
helpful discussions. We gratefully acknowledge support by the
german science foundation DFG under Grant Nr.~BA~1477-12-2.


\end{document}